\documentclass[prl,twocolumn,showpacs,superscriptaddress,groupedaddress]{revtex4}

\usepackage{amssymb,amsfonts,amsmath} 

\usepackage{graphicx} 

\usepackage{float}

\begin{document}
\title{Expansion dynamics in the one-dimensional Fermi-Hubbard model} 

%\preprint{}



%\author{}
%\thanks{}
%\affiliation{}

\author{J.\ Kajala} 
\affiliation{Department of Applied Physics,  
Aalto University School of Science, P.O.Box 15100, FI-00076 Aalto, FINLAND} 
\author{F.\ Massel}
\affiliation{Low Temperature Laboratory,
 Aalto University, P.O. Box 15100, FI-00076 Aalto, FINLAND} 
\author{P. T\"orm\"a}
\email{paivi.torma@hut.fi}
\affiliation{Department of Applied Physics,  
Aalto University School of Science, P.O.Box 15100, FI-00076 Aalto, FINLAND}
\affiliation{Kavli Institute for Theoretical Physics, University of California, Santa Barbara, California 93106-4030, USA}


\date{January 31, 2010}

\begin{abstract}

  Expansion dynamics of interacting fermions in a lattice are
  simulated within the one-dimensional (1D) Hubbard model, using the
  essentially exact time-evolving block decimation (TEBD) method. In
  particular, the expansion of an initial band-insulator state is
  considered. We analyze the simulation results based on the dynamics
  of a two-site two-particle system, the so-called Hubbard dimer. Our
  findings describe essential features of a recent experiment on the
  expansion of a Fermi gas in a two-dimensional lattice. We show that the
  Hubbard-dimer dynamics, combined with a two-fluid model for the
  paired and non-paired components of the gas, gives an efficient
  description of the full dynamics. This should be useful for
  describing dynamical phenomena of strongly interacting Fermions in
  a lattice in general.
\end{abstract}

%67.85.-d, 74.50.+r, 03.75.Lm}

% THE PACS NOS NEED TO BE CHANGED *****************************************************

%\keywords{}

\maketitle

Important physical phenomena such as magnetism and high-temperature
superconductivity are often approached by theories based on the
Hubbard model \cite{HLEssler:2005p1368,LeHur:2009p331} which describes
interacting particles in a lattice. Within ultracold gas systems
\cite{Jaksch:2005p1471,Bloch:2005p1473}, the Hubbard model can be
efficiently realized and studied in experiments with bosonic
\cite{Greiner:2002p1470} and recently with fermionic atoms
\cite{Joerdens:2008p276,Schneider:2008p1474}. Intriguingly, the
dimension can be easily controlled. Low-dimensional systems such as
nanowires, iron pnictides and graphene are currently highlighted
topics of research. Models for the quantum many-body physics of 2D and
1D systems can explored with ultracold gases, c.f. recent experiments
on fermions in one dimension \cite{Liao:2010p1428} and
expanding fermions in a 2D lattice \cite{Schneider:2010p1468}. For
one-dimensional systems, an advantage is that the experiments can be
compared to exact theoretical descriptions. However, although
the ground state and static properties of one-dimensional systems are
known to an impressive degree
\cite{HLEssler:2005p1368,Giamarchi:2004p1570}, {\it dynamics} is
largely unexplored. Work on theory and simulation of dynamical
properties of interacting fermions in 1D has recently been emerging
\cite{Kollath:2005p1574}.

In this Letter, we study with exact numerical methods the expansion of
fermions within the one-dimensional Hubbard model. We show that the
resulting complex dynamics can be efficiently described by a
two-fluid model in which we deduce the dynamics of the fluids from the
dynamics of a Hubbard dimer. Our results explain several main features
of the experiment \cite{Schneider:2010p1468} performed in 2D, and
give exact predictions for future experiments in 1D. The simple
Hubbard-dimer two-fluid model that we have developed provides a basis
for the description of various types of expansion, collision and
oscillation dynamics for fermions in lattices.

\begin{figure}
  \includegraphics[width=0.45\textwidth]{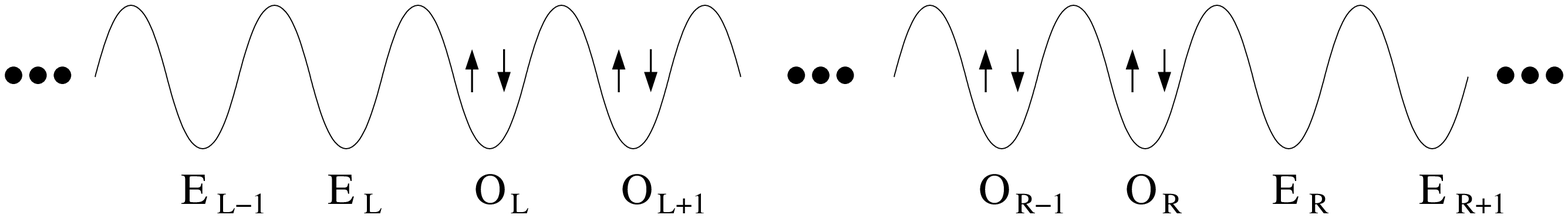} 
  \caption {(color online) Schematic representation of the initial state: the middle
  part of the lattice is fully occupied ($O_i$) and the rest is empty
  ($E_i$). Sites $E_L$, $O_L$ and $O_R$, $E_R$ represent the left
  and right edge of the cloud, respectively. }
\label{fig:sch} 
\end{figure}

We use the time-evolving block decimation (TEBD) algorithm
\cite{Vidal:2003p622} to describe
the time evolution generated by the Hubbard Hamiltonian
%\begin{align}
%\label{eq:HH} 
 $  H_H=U \sum_i \hat{n}_{i,\uparrow}\hat{n}_{i,\downarrow} 
- J \sum_{i\,\sigma=\uparrow,\downarrow} c^\dagger_{i\,\sigma}c_{i+1\,\sigma} + h.c. $,
%\end{align}
given an initial state $|\phi(0)\rangle$, where
$\hat{n}_{i,\sigma}=c^\dagger_{i\,\sigma}c_{i\,\sigma}$ with
$c^\dagger_{i\,\sigma}$ and $c_{i\,\sigma}$ representing the creation
and annihilation of a fermion with spin $\sigma$ at the site $i=1\dots
L$. Moreover, the initial state is given by
$|\phi(0)\rangle=|\emptyset\rangle_1 \dots
|\emptyset\rangle_{E_L}|\uparrow\,\downarrow\rangle_{O_L}\dots|\uparrow\,\downarrow\rangle_{O_R}
|\emptyset\rangle_{E_R} \dots |\emptyset\rangle_{L}$ (see Fig. 
\ref{fig:sch}).
The initial state consists thus of a band insulator occupying the
central $O_L-O_R$ sites of an otherwise empty lattice. In the
simulation we have considered $L=150$, $E_L=66$, $O_L=67$, $O_R=86$,
$E_R=87$, the Schmidt number $\xi=150$, and $J = 1$.
% In the experiment \ref{Schneider:2010p1468} the band insulator ground state was
% created by having a very large lattice height, effectively trapping
% the pairs in single lattice sites. Dynamics was then initiated by
% lowering the lattice height.  However, numerics is somewhat easier
% because we can just input the band insulator ground state as an input
% to the code, and then subject this ground state to the TEBD time
% evolution \cite{Daley:2004p106, Leskinen1} under the Hubbard Hamiltonian
% \eqref{eq:HH}.  The band insulating ground state is such that we have
% the state $|\uparrow \downarrow\rangle$ in lattice sites 67-86 and the empty
% state $|\phi\rangle$ elsewhere. We need to note here that we do the
% simulation at zero temperature, as the ground state input by hand does
% not have thermal excitations and the subsequent TEBD dynamics is done
% at zero temperature.  Therefore the numerics has two differences to
% the experiment: the dimensionality and the absence of thermal
% excitations. Nonetheless, thermal excitations should not play a major
% role in the expansion dynamics.
 Our code allows us to access the expectation value of the
 (spin-resolved) local particle number $n_{i\,\uparrow}(t)$ and
 $n_{i\,\downarrow }(t)$ along with the local double occupancy
 %\begin{align} 
 %  \label{eq:PairCorrelator} 
  $   n_{i\, \uparrow \downarrow} (t) = \langle \phi (t) |\hat{n}_{i\, \uparrow} \hat{n}_{i\, \downarrow} |\phi (t)  \rangle. $
 % \end{align}
  Note that $n_{i\, \downarrow} (t)=n_{i\, \uparrow} (t)$ since the
  problem is spin symmetric. In our analysis we will show that the
  evolution of the initial state can be described in terms of a
  two-fluid model where the two fluids are represented by single
  particles and \textit{doublons} as suggested by the Bethe-ansatz
  solution of the Hubbard model \cite{HLEssler:2005p1368} and has been
  shown in the context of imbalanced Fermi gases
  \cite{Zhao:2010p1469,Orso:2007p1758,Hu:2007p1836,Hu:2007p1864,Guan:2007p1985}.
  The doublons are excitations of the form
  $c_{i\,\uparrow}^\dagger c_{i\, \downarrow}^\dagger|\emptyset
  \rangle$ and the single (unpaired) particles are defined as
  $c_{i\,\sigma}^\dagger|\emptyset\rangle$
  ($\sigma=\uparrow,\,\downarrow$). The local number of doublons is
  given by $n_{i\, \uparrow \downarrow} (t)$, while the number of unpaired
  (up) particles is given by  $n_{i\, \uparrow}^{un} (t) =  n_{i\, \uparrow} (t) - n_{i\,\uparrow \downarrow} (t)$. 
  %$n_{i\,\uparrow}-n_{i\, \uparrow %    \downarrow}$.   
  
%  Another quantity that will be relevant to our analysis is the
 % fraction of unpaired (up) particles 

\begin{figure}
\includegraphics[width=0.45\textwidth]{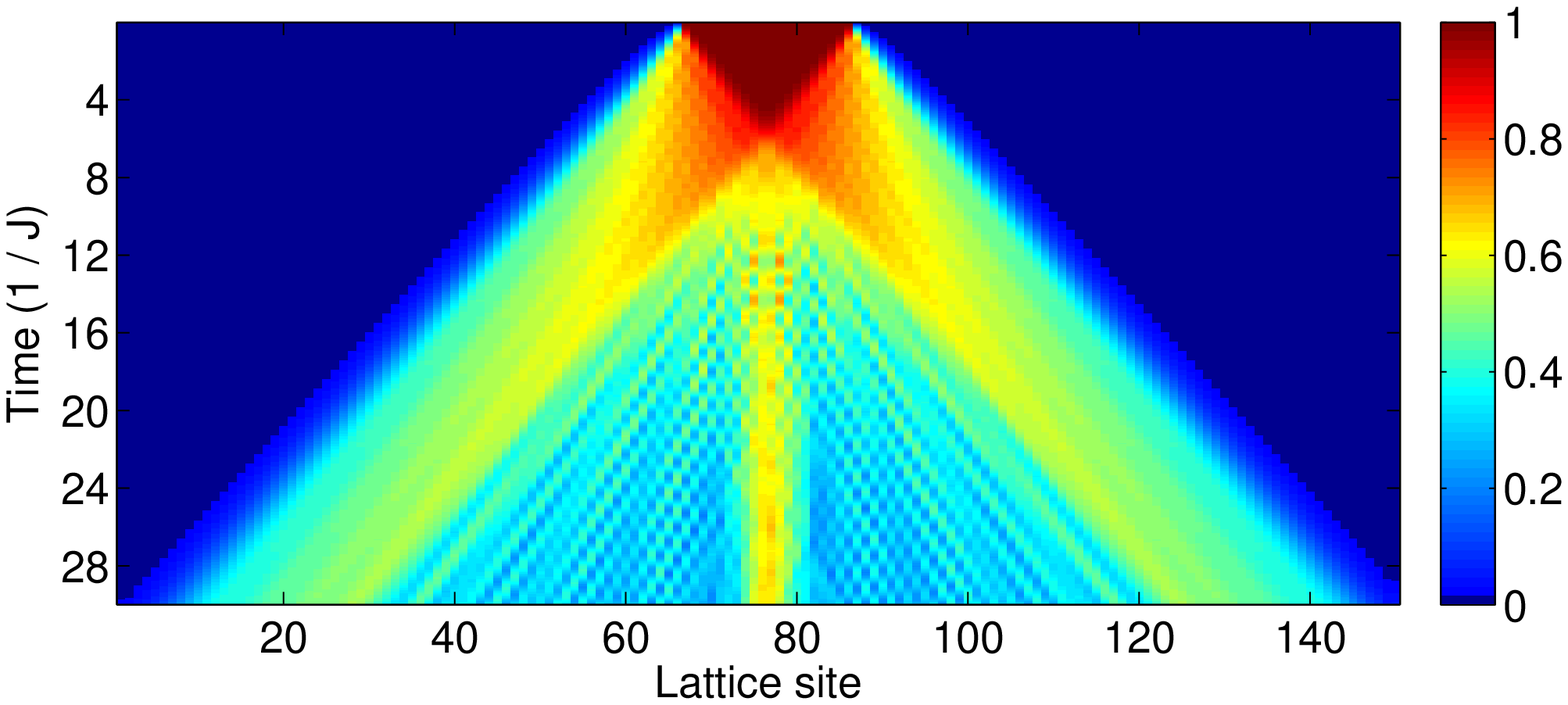} 
\includegraphics[width=0.45\textwidth]{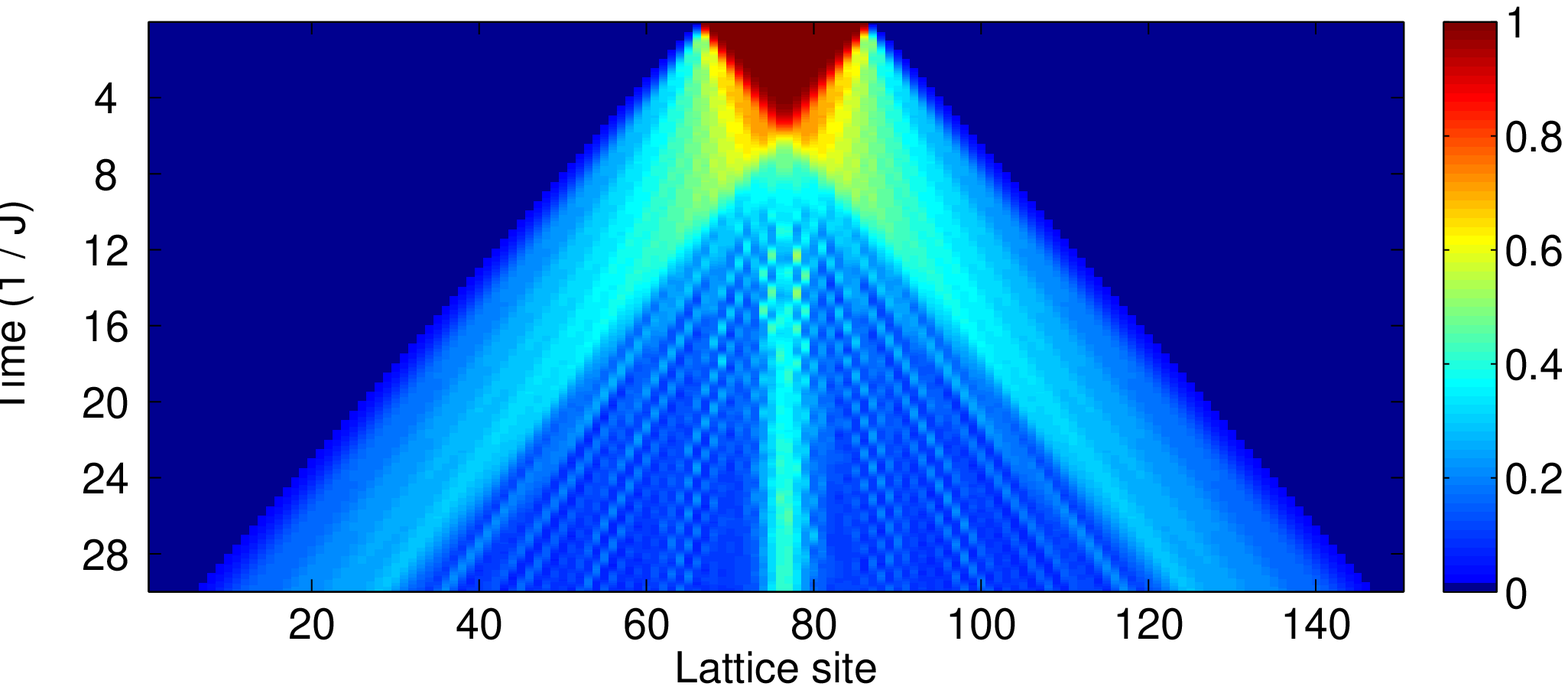} 
\caption {(color online) Time evolution of $\sqrt{n_{i\, \uparrow}
    (t)}$ (above) and $\sqrt{n_{i\, \uparrow \downarrow} (t)}$ (below)
  for $|U|=0.0$. The free-particle nature of the expansion is clear
  from the absence of separate doublon expansion wavefronts.}
\label{fig:U0.0} 
\end{figure}

\begin{figure}
\includegraphics[width=0.45\textwidth]{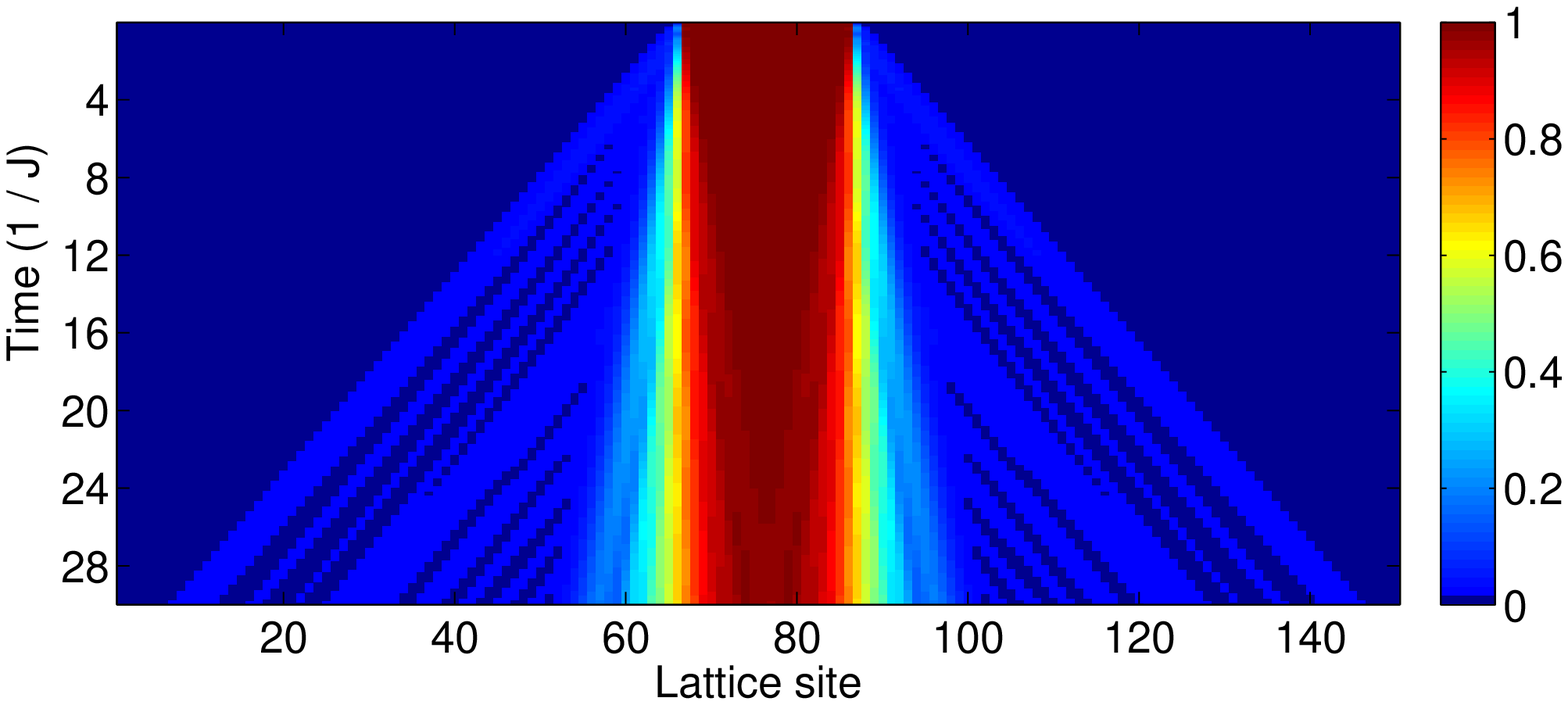} 
\includegraphics[width=0.45\textwidth]{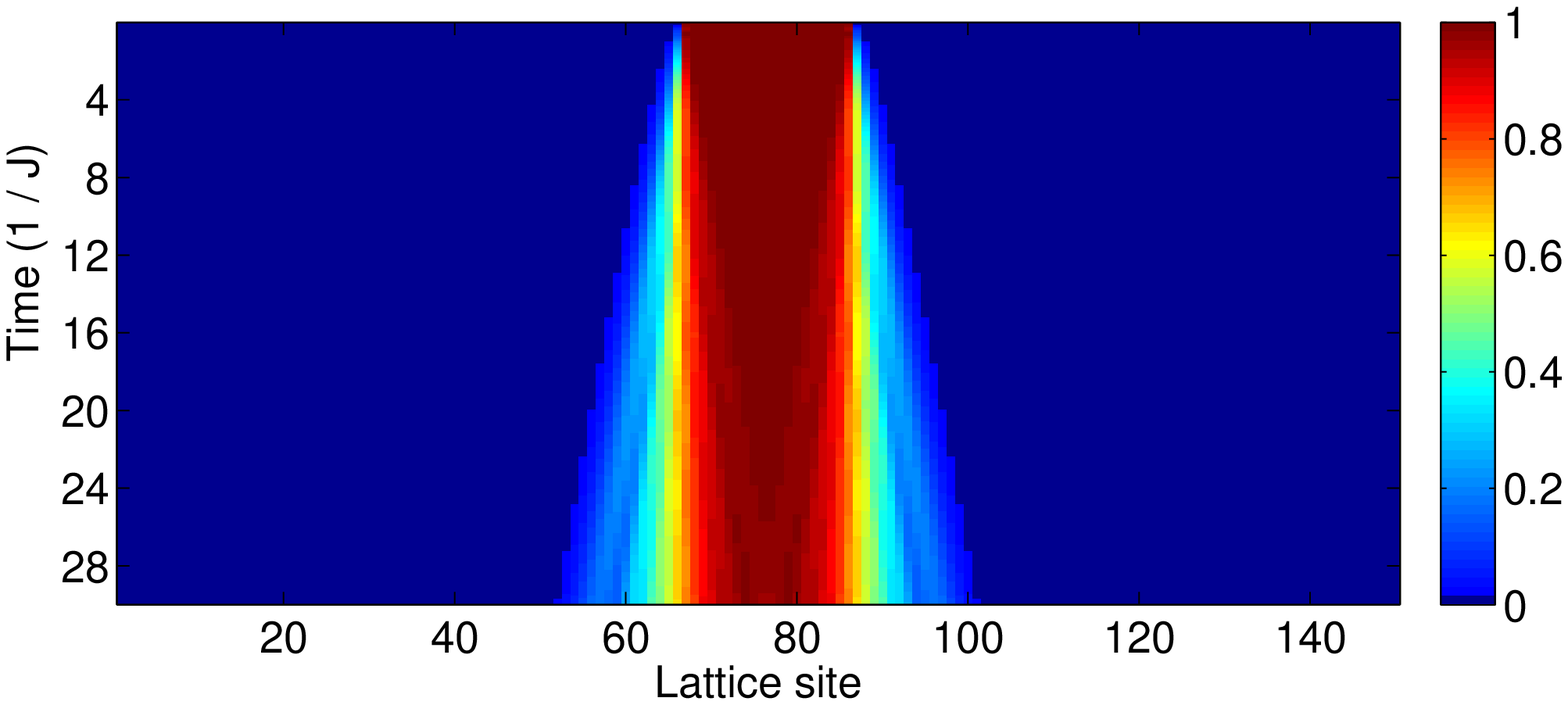}
\caption { (color online) Same as Fig. \ref{fig:U0.0} for
  $|U|=10.0$. For this interaction it is possible to distinguish the
  two wavefronts.}
\label{fig:U10.0} 
\end{figure}

  In  Figs. \ref{fig:U0.0} and \ref{fig:U10.0}, $\sqrt{n_{i\, \uparrow} (t)}$ 
  and $\sqrt{n_{i\, \uparrow \downarrow} (t)}$ are depicted
  for $U = 0.0$ and $U = \pm 10.0$.  We are plotting the square roots
  of the densities since this highlights low density features (see the
  supplementary material for the full density plots). As in general
  for the spin-balanced Hubbard model \cite{HLEssler:2005p1368}, the
  density distributions evolve in time exactly in the same way for
  $U$ and $-U$. This $U \leftrightarrow -U$ symmetry holds also for
  all observables for all interactions and it was indeed observed also in
  the experiment \cite{Schneider:2010p1468}.

% \begin{figure}
% \includegraphics[width=0.45\textwidth]{nupU+1.0.eps} 
% \includegraphics[width=0.45\textwidth]{nupdownU+1.0.eps}
% \caption 
% { 
%  $\sqrt{n_{i\, \uparrow} (t)}$ (left) and $\sqrt{n_{i\, \uparrow \downarrow}
%    (t)}$ (right) for $|U|=1.0$.
% } 
% \label{fig:U1.0} 
% \end{figure}  

In the non-interacting case in Fig. \ref{fig:U0.0}, both
particles and doublons are expanding at the speed of $2J$, corresponding to the
highest group velocity allowed by the dispersion relation (see supplementary material). 
%The same behaviour is observed in
%the case of interaction $|U| = 0.1$.
%
% Is it just that actually we can't tell the difference? 
% Saying this here might be a bit conceptually confusing... 
For strong interactions (Fig. \ref{fig:U10.0}), we see two separate
wavefronts. Such separation into two types of wavefronts is clearly
observable for interactions $|U| > 3.0$.  The outermost wavefront
consists of fully unpaired particles expanding at the of speed $2J$, like in
the $U=0$ case. In contrast \textit{doublons} expand
at the speed of $4J^2/U$ (see supplementary material for the explanation of
these results).

%%%%%%%%%% TO SUPPLEMENTARY %%%%%%%%%%%%%%%%%%%%%%%%%%%%%%%%%%%%%%%%%%%
% The value for the expansion speed can be obtained by mapping
% the Hamiltonian for the doublons hopping through an empty lattice to
% an isotropic Heisenberg Hamiltonian and, subsequently, to a spinless
% fermions Hamiltonian with nearest neighbor interaction
% \cite{Giamarchi:2004p1570} (see supplementary material). The structure
% of the initial state, however, preventing hopping of doublons towards
% the center of the cloud, allows to neglect the nearest-neighbor
% interaction.

\begin{figure}
\includegraphics[width=0.45\textwidth]{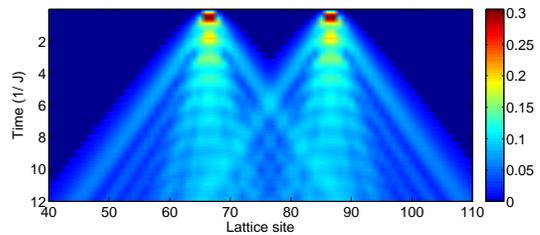} 
\caption { (color online) Unpaired particle expansion $n_{i\,
    \uparrow}^{un} (t)$ for $|U| = 5.0$, exhibiting the oscillations
  described in the text.}
\label{fig:unpaired} 
\end{figure}

\begin{figure}
\includegraphics[width=0.45\textwidth]{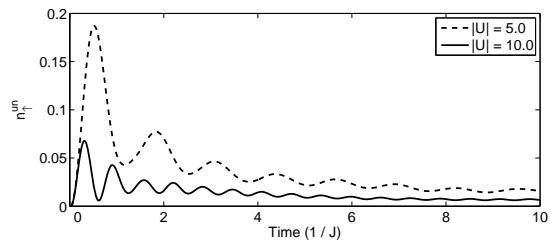} 
\caption { Unpaired population dynamics $n_{E_L + O_L , \uparrow}^{un}(t)$
  for $|U| = 5.0$ and $|U| = 10.0$.  }
\label{fig:6667unpaired} 
\end{figure}

Intermediate interactions $0.5 \leq |U| \leq 3.0$ show a more
complicated behaviour. The separate expansion fronts are no longer
well distinguishable, suggesting a stronger interplay between single
particles and doublons. Moving to even lower interactions, the
unpaired particles and the doublons behave similarly to the case of
$U=0$. Let us now examine the time dependence of the number of unpaired
(up) particles $n_{i\, \uparrow}^{un} (t)$ for $|U| = 5.0$ (Fig.
\ref{fig:unpaired}).
Initially the dynamics in the band insulator cloud is Pauli blocked,
since neighbouring lattice sites in the center have unit density for
both spin up and down. Therefore the unpaired expansion fronts are
created at the edges of the cloud.
Intriguingly, $n_{i\, \uparrow}^{un} (t)$ shows
damped oscillations at the edges associated with the emission of
unpaired particles into the empty lattice. Considering the time
evolution of $n_{E_L, \uparrow}^{un} + n_{O_L, \uparrow}^{un}$ (the two edge sites) over the whole interaction range
$|U| = 0.0 - 15.0$ (see Fig. \ref{fig:6667unpaired} for $|U|=5.0,\,
10$) we find that there are fewer periods of oscillations for lower
interactions (for $|U| = 1.0$ only one broad oscillation peak is
visible) and the oscillation frequency increases with 
interaction strength, see 
supplementary material for a general survey of the data.

\begin{figure}
\includegraphics[width=0.45\textwidth]{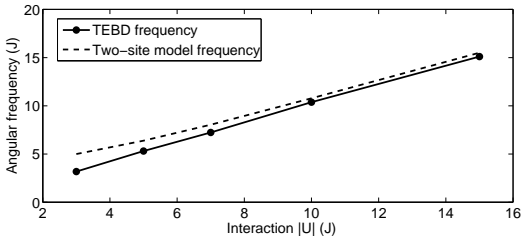} 
\caption 
{The frequency given by the FT of $n_{E_{L} + O_{L}, \uparrow}^{un} (t)$
(from TEBD numerics) compared to the frequency $\sqrt{U^2 + 16J^2}$ obtained by solving the two-site model.} 
\label{fig:freqcomp} 
\end{figure}

\begin{figure}
\includegraphics[width=0.4\textwidth]{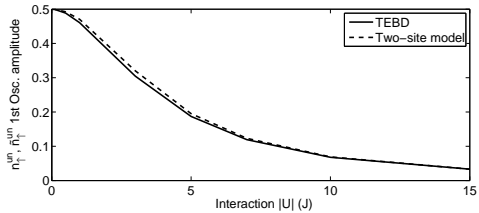} 
\caption 
{ The height of the first peak of $n_{E_{L} + O_{L}, \uparrow}^{un} (t)$ (from TEBD numerics) 
  compared to the amplitude obtained by solving the two-site model.} 
\label{fig:amplitudecomp} 
\end{figure}

The evolution of doublons into unpaired particles plays a key role in the
expansion physics. For this reason, we focus now on the explanation of the
oscillations in the case of high interactions, see
Figs. \ref{fig:unpaired} and \ref{fig:6667unpaired}. Our hypothesis is
that one can consider the edges of the cloud (sites $E_L$, $O_L$ and
$O_R$, $E_R$) at short timescales as two-site systems (Hubbard
dimers \cite{HLEssler:2005p1368, Giamarchi:2004p1570}). Focusing on the $E_L$/$O_L$ dimer, the system can be
described as an initially empty state $|\emptyset\rangle$ in the left
lattice site $E_L$ and a doublon $|\uparrow \downarrow\rangle$ in the
right lattice site $O_L$. The dynamics of the dimer with this
initial state can be solved analytically by diagonalizing its
Hamiltonian (see supplementary material). As a result, in the two-site
problem, the population of unpaired up particles on the two sites $E_L$ and $O_L$ 
is given by (the tilde refers to the Hubbard dimer model):

\begin{align} 
  \label{eq:n_unpaired_analytic} 
  \tilde{n}_{E_L + O_L, \uparrow}^{un} (t) = 
          \frac{1 - \cos(\sqrt{U^2 + 16J^2}t)}
          {2 + \frac{U^2}{8J^2}}.
\end{align} 
We extract the oscillation frequencies
from the numerical Fourier transform (FT) of $n_{E_L+O_L,
  \uparrow}^{un}$ and compare its peaks to the oscillation frequency
$\sqrt{U^2 + 16J^2}$ in \eqref{eq:n_unpaired_analytic} (see Fig. \ref{fig:freqcomp}). Moreover, we
compare the height of the first peak of the unpaired density
oscillations (seen in Figs. \ref{fig:unpaired} and \ref{fig:6667unpaired}) to the amplitude
$8/\left(16 + \frac{U^2}{J^2}\right)$, in Fig. \ref{fig:amplitudecomp}.

The agreement is good considering that, for longer times,
the FT of $n_{E_L + O_L, \uparrow}^{un} (t)$ has additional contributions
stemming from the hopping between the dimer and the rest of
the chain. However, as the frequency is approximately $U$ in the high interaction limit, 
one might claim that the frequency correspondence could be obtained from simple energy arguments. 
Therefore, to further confirm the validity of the Hubbard Dimer model, we now move on
to examine whether the
two-site model coupled to the next adjacent sites explains the decay
observed in $n_{E_L + O_L, \uparrow}^{un} (t)$ (see
Fig. \ref{fig:6667unpaired}).
Let us define the damping $D$ to mean the decrease of the amplitude of the
unpaired density oscillations, compared to $t=0$. We propose that the
damping should be equal to the probability of having an unpaired
particle in the Hubbard dimer times the probability for this single particle
tunnelling out of this system. The probability for the single particle
tunnelling (obtained by solving the two-site system with the
initial state $|\emptyset, \uparrow\rangle$) is given by
$\sin^2(J\,t)$. Combining this result with
Eq.\eqref{eq:n_unpaired_analytic}, we obtain the
damping at a given time $\tau$:
\begin{align} 
  \label{eq:damping} 
  D(\tau) = 2 \int_{0}^{ \tau} 
        \frac{ 1 - \cos(\sqrt{U^2 + 16J^2}t)}
             {2 + \frac{U^2}{8J^2}}  
        \sin^{2}(J\,t) dt ,
\end{align} 
where the factor of two comes from the particle-hole symmetry:
particles leaking out from the dimer to the left (to $E_L-1$) are
mirrored by holes leaking to the right (to $O_L+1$) thus generating
%%%%%%%%%% TO SUPPLEMENTARY %%%%%%%%%%%%%%%%%%%%%%%%%
% Formally, we can say that the partial trace $Tr_{L\setminus
%   E_L-1}\left\{|\phi(t)\rangle\langle\phi(t)|\right\}$ translates to a
% (non-unitary) evolution of the state $|\emptyset\rangle_{E_L-1}$ into
% a superposition of $|\uparrow\rangle_{E_L-1}$ and
% $|\downarrow\rangle_{E_L-1}$, effectively increasing the particle
% number, the state $|\uparrow \downarrow\rangle_{O_L+1}$ evolves to the
% superposition of $|\uparrow\rangle_{O_L+1}$ and
% $|\downarrow\rangle_{O_L+1}$, leading to a decrease of the particle
% number in the site $O_L+1$.
%%%%%%%%%%%%%%%%%%%%%%%%%%%%%%%%%%%
particle-like expansion fronts emitted out of the initial cloud and
hole-like expansion fronts emitted into the cloud. When the particle-like
and hole-like expansion fronts meet interference 
in the unpaired particle density is visible, see Fig. \ref{fig:unpaired}. 

%The merging of the
%unpaired particles decaying around the edge $E_L/O_L$, with the
%unpaired particles incoming from the opposite edge
%(Fig. \ref{fig:unpaired}), causes an increase of $n_{E_L\,
%  O_L}^{un}(t)$ at time $10J^{-1}$.
% We note that in the numerics we indeed observe particle-hole
% symmetry for the initial time developement.  ent. This means that we
% have $n^{66}_{\uparrow \downarrow} (t) = 1 - n^{67}_{\uparrow
%   \downarrow} (t)$, $n^{65}_{\uparrow \downarrow} (t) = 1 -
% n^{68}_{\uparrow \downarrow} (t)$ , ... and $n^{66}_{\uparrow,
%   unpaired} (t) = n^{67}_{\uparrow, unpaired} (t)$,
% $n^{65}_{\uparrow, Singlet} (t) = n^{68}_{\uparrow, Singlet} (t)$,
% ... .  The symmetry is however eventually broken by the finite size
% of the band insulator cloud.

\begin{figure}
\includegraphics[width=0.45\textwidth]{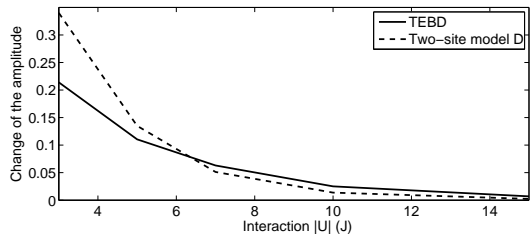} 
\caption {Change of the amplitude of $\tilde{n}_{E_L, \uparrow}^{un} (t) $
  oscillations after $t = \frac{3 \pi}{\sqrt{U^2 + 16J^2}}$, compared
  to the change of amplitude between the first and second oscillation
  peaks of  $n_{O_L + E_L, \uparrow}^{un} (t) $ observed in TEBD
  numerics.}
\label{fig:dampingcomp} 
\end{figure}

By comparing the decay predicted in
Eq. \eqref{eq:damping} to the numerics, we observe that, for the duration
of the first half-period, the decay is negligible. This is in accordance with the
height of the first peak being equal to the two-site oscillation
amplitude, as shown in Fig.  \ref{fig:amplitudecomp}. After three half-periods we compare
the damping as predicted by Eq. \eqref{eq:damping} to the change of amplitude between
the first peak and the second peak seen in numerics, see Fig. \ref{fig:dampingcomp}.
The two-site model is again in good agreement with TEBD for $|U| > 3.0$. 
The time beyond which Eq. \eqref{eq:damping} fails to describe the
expansion physics is when the population of unpaired particles in the
sites $O_{L+1}, E_{L-1}$ becomes significant. This occurs when the
$sin^2(Jt) = \frac{1 - 2 cos(2Jt)}{2}$ term is no longer close to zero,
limiting our short time analysis to times
$t << \frac{\pi}{2} \frac{1}{J}$.
When U is sufficiently large, the Hubbard dimer oscillations occur in much shorter timescale than the single particle tunneling does. In other words, a large number of oscillations occur in the window $0 < t < \frac{\pi}{2} \frac{1}{J}$
In the case of lower interactions, Hubbard Dimer oscillations at frequency
$\sqrt{U^2 + 16J^2}$ become comparable to the frequency $2J$ and therefore we see that already the first oscillation peaks
are heavily damped.

%lower interactios no peaks in FT well visible

%For lower interactions our
%assumptions break down, as, in the neighbouring sites, the density
%deviations from the initial values become relevant already for short
%times.
% Moreover, comparing $D$ at $\tau = \frac{5 \pi}{\sqrt(U^2 + 16J^2}$
% to the height of the third peak seen in the numerics shows that the
% short time approximation breaks also for all but the highest
% interactions ($U = 15.0$).

%describe the expansion physics agrees with our analytical
%predictions, corresponding to the conditions $t << \pi, t << U * \pi / 2$ which is when
%the Hubbard dimer oscillations occur in much shorter timescale than single particle and pair tunneling, respectively.  
%At this time the unpaired particle populations of the
%adjacent sites become comparable to that of the dimer, in agreement
%with the numerical data.
% We would obtain better match also for longer times if we included in
% the analysis 1) updating the unpaired amplitude by the density that
% has leaked out of the system 2) including unpaired particles coming
% back to the system from outside of the system. But for longer times,
% this would involve adding first sites 65 and 68 to the analysis,
% then 64 and 69 and so on, making the analytics more complicated by
% each addition.

%Let us now reflect on what the results we obtained mean. 
In general, the two-site dynamics is well able to describe the creation
of the particle, hole, and doublon wavefronts seen in the density
profiles. These wavefronts are created during the two-site
oscillations. 
% Considering the non-interacting limit, the paired state and unpaired
% state are degenerate, And we can think the paired and unpaired states
% as a two level system with equal energies.  In such a system the
% density expectation value in each state is 0.5 (with initially 1.0 in
% the paired state and 0.0 in the unpaired state).  Particles in both
% states will behave like non-interacting particles. This is in good
% agreement with the low interaction density profiles obtained from
% numerics: the paired and unpaired densities become quickly equal at
% the edges, then both expand in equal density wavefronts with speed
% $2J$.
% In both the low and high interaction limit, therefore, it seems that
% we can describe the system with a two-fluid model.
Our numerical results and analytical investigations confirm the
two-fluid picture of the system. Initially, the interaction takes
place at the edges of the cloud, where unpaired particles are created
according to the dimer dynamics described above. The subsequent
expansion is explained by dynamics of non-interacting particles (at
the speed of $2J$) or doublons (at the speed of $4J^2/U$).  Our model gives an
excellent quantitative description in the highly-interacting limit due
to the clear separation of the expansion and dimer oscillation
eigenfrequencies. For interactions $0.5 \leq |U| \leq 3.0$
the interplay between the expansion and the Hubbard dimer dynamics does not
allow a quantitative description of the numerical results, however it
provides a qualitative framework for further analysis. For $|U|\leq
0.5$, the free-particle expansion seems to give a fairly good
description.

Finally, we compare our results to the 2D experiment of
\cite{Schneider:2010p1468} done at finite temperature. We suggest that our two-site
considerations also apply to the dynamics of the experiment. In 2D
there is coupling to four adjacent sites, but just like in 1D,
initially the sites are either Pauli blocked or empty. The simplified
dynamics should originate from the two-site analysis, and subsequent
short-time dynamics in the high interaction limit correspond to the
two-fluid expansion, with the two fluids interacting only at the
edges. 

%Making the connection with to the classical non-linear diffusion model in \cite{Schneider:2010p1468} we have
%in the high interaction limit ballistic particles emitted from the 
%diffusive core, the two types of ballistic particles being unpaired particles
%and doublons. These two expand at constant speeds ($2J$ and $\frac{4J^2}{U^2}$)
%in wavefronts whose densities are so low that spin-spin scattering 
%(i.e. the Hubbard Dimer dynamics) is negligibile.  In contrast, the core has high density, 
%the Dimer dynamics occur, and thus the core is diffusive. In the case of interactions lower than $|U| < 3$ 
% also the densities of the unpaired and doublon wavefronts become high enough for the Dimer Dynamics, 
%which in the language of  the model used in \cite{Schneider:2010p1468} makes the two types of 
%wavefronts diffusive. Finally, for the very low interactions $U \leq 0.5$ 
%the problem simplifies to ballistic expansion of non-interacting particles. 

\begin{figure}
\includegraphics[width=0.45\textwidth]{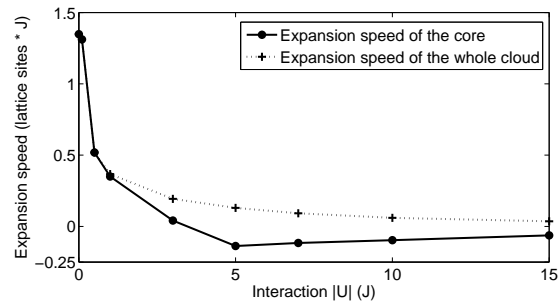} 
\caption { The core expansion speed $\dot{R}(t)$, as a function of the
  interaction strength. For reference, we plot also the expansion speed
of the whole cloud.}
\label{fig:experimentcomp} 
\end{figure}

In order to compare our results to Fig. 5 of \cite{Schneider:2010p1468}
we define the core density as $n_{i\, \uparrow}^{C} (t) = 
n_{i\, \uparrow} (t) - n_{i\, \uparrow}^{b} (t)$ for $|U| > 0.5$ and
$n_{i\, \uparrow}^{C} (t) = n_{i\, \uparrow} (t)$  for  $|U| \leq 0.5$,
where $n_{i\, \uparrow}^{b} (t)$ is the density of purely ballistic particles (see supplementary material). The definition of the core then corresponds to
the diffusive part of the cloud in the model of \cite{Schneider:2010p1468}. 
The core expansion velocity is given by $\dot{R}(t)$, where
$R(t) = \sqrt{<i^2> - <i>^2}$, 
$<i> = \sum_{i=1}^{L}([n_{i\, \uparrow}^{C} (t) + n_{i\, \downarrow}^{C} (t)] * i) 
/ \sum_{i=1}^{L}(n_{i\, \uparrow}^{C} (t) + n_{i\, \downarrow}^{C} (t))$ 
and $L$ is the number of lattice sites. 
The core expansion speed as a function of interaction is plotted in Fig. \ref{fig:experimentcomp}.
The behavior of $\dot{R}(t)$ is indeed similar to the core expansion 
velocity in Fig. 5 of \cite{Schneider:2010p1468}, showing also the negative velocities.

Another recent experiment \cite{Sommer:2010p1468} studied collision
dynamics of two Fermi gas clouds. Although the experiment is not done in a lattice,
the theoretical framework presented here can in the low density limit
be used to desribe also physics in \cite{Sommer:2010p1468},
c.f. \cite{Kajala:2011}. 

In conclusion, we studied the expansion of an interacting fermionic gas
in a 1D lattice. We showed that the time evolution of this system can
be described in terms of a two-fluid model of unpaired particles and
doublons whose interplay gives rise to nontrivial dynamics. We 
suggest that the experimental results of \cite{Schneider:2010p1468}
can be interpreted in terms of the analysis performed here. 
Our results should be widely applicable since expansion is a basic 
dynamics problem related to the ultracold gas experiments in particular,  
and to the transport properties of fermions in general.

\begin{acknowledgments} 
  We thank I. Bloch and U. Schneider for useful discussions. This work
  was supported by the Academy of Finland (Projects No. 213362,
  No. 217043, No. 217045, No. 210953, and No. 135000) and
  EuroQUAM/FerMix, and conducted as a part of a EURYI scheme grant
  (see www.esf.org/euryi). The research was partly supported by
  the National Science Foundation under Grant No. PHY05-51164.
  Computing resources were provided by CSC –
  the Finnish IT Centre for Science.
\end{acknowledgments} 

%%\bibliographystyle{apsrev_abb}
%%\bibliography{BI_bib}

\section{Expansion dynamics in the one-dimensional Fermi-Hubbard model - Supplementary material}

\section{Survey of the numerical results}
\label{sec:num_res}

\subsection{Density profiles}
\label{sec:dens_prof}

\begin{figure}[!h]
  \includegraphics[width=0.45\textwidth]{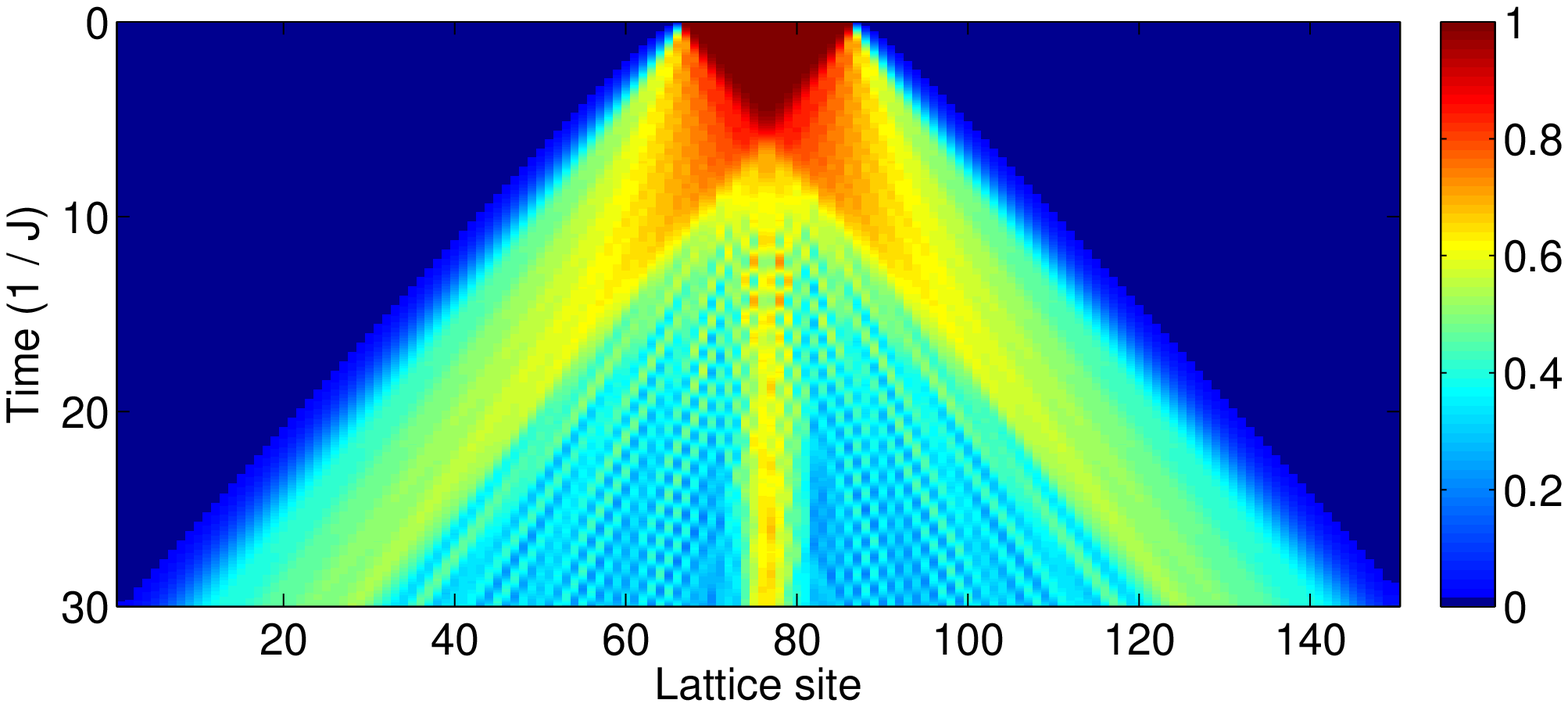}
	\includegraphics[width=0.45\textwidth]{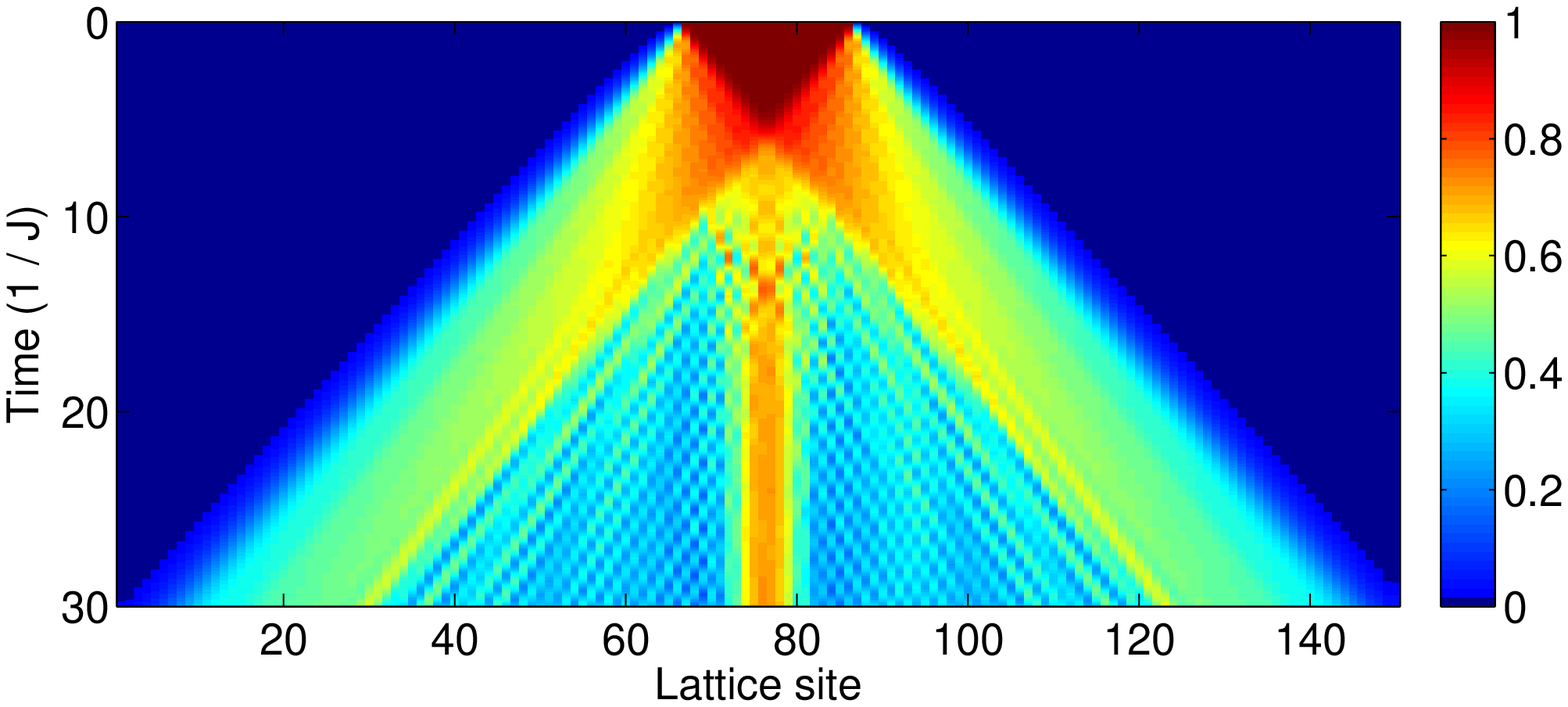}
        \includegraphics[width=0.45\textwidth]{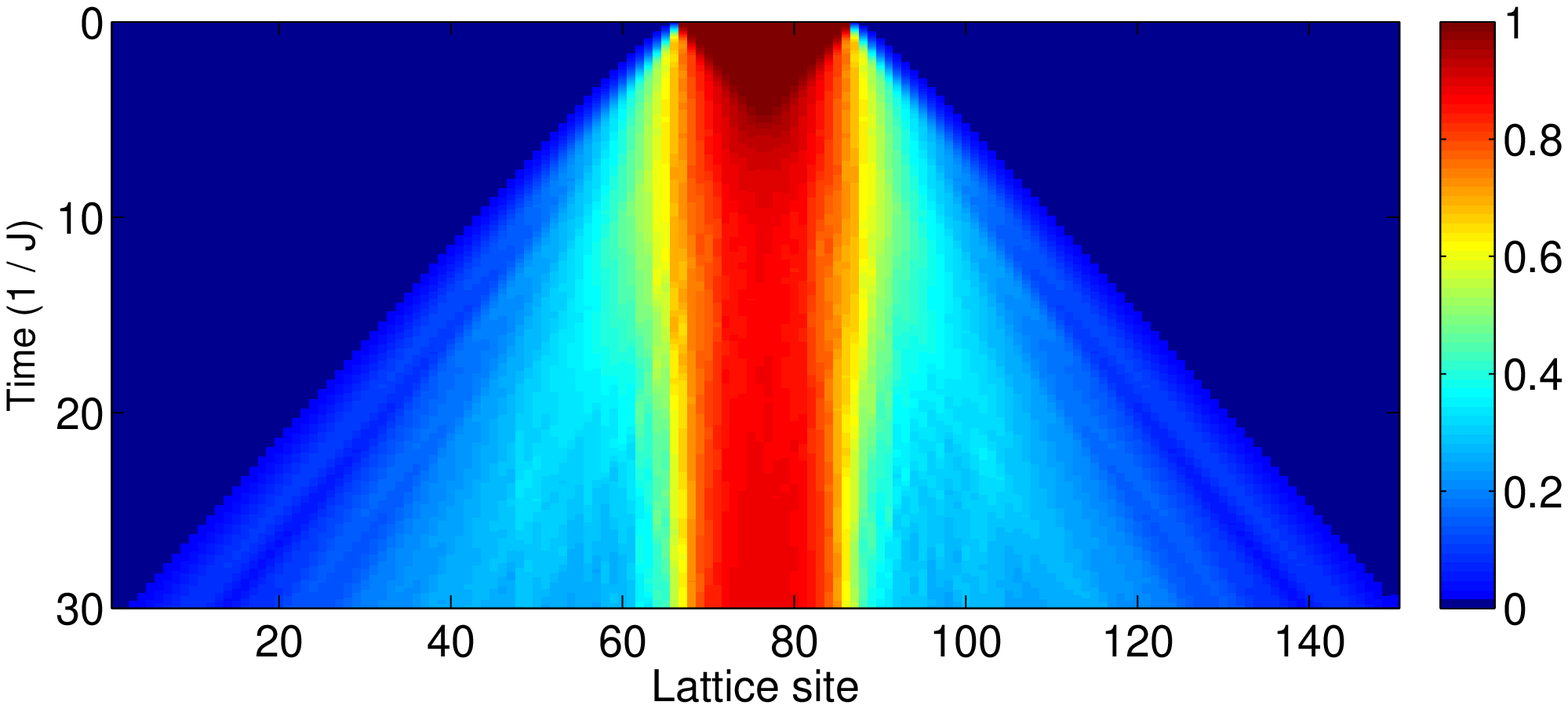}
        \includegraphics[width=0.45\textwidth]{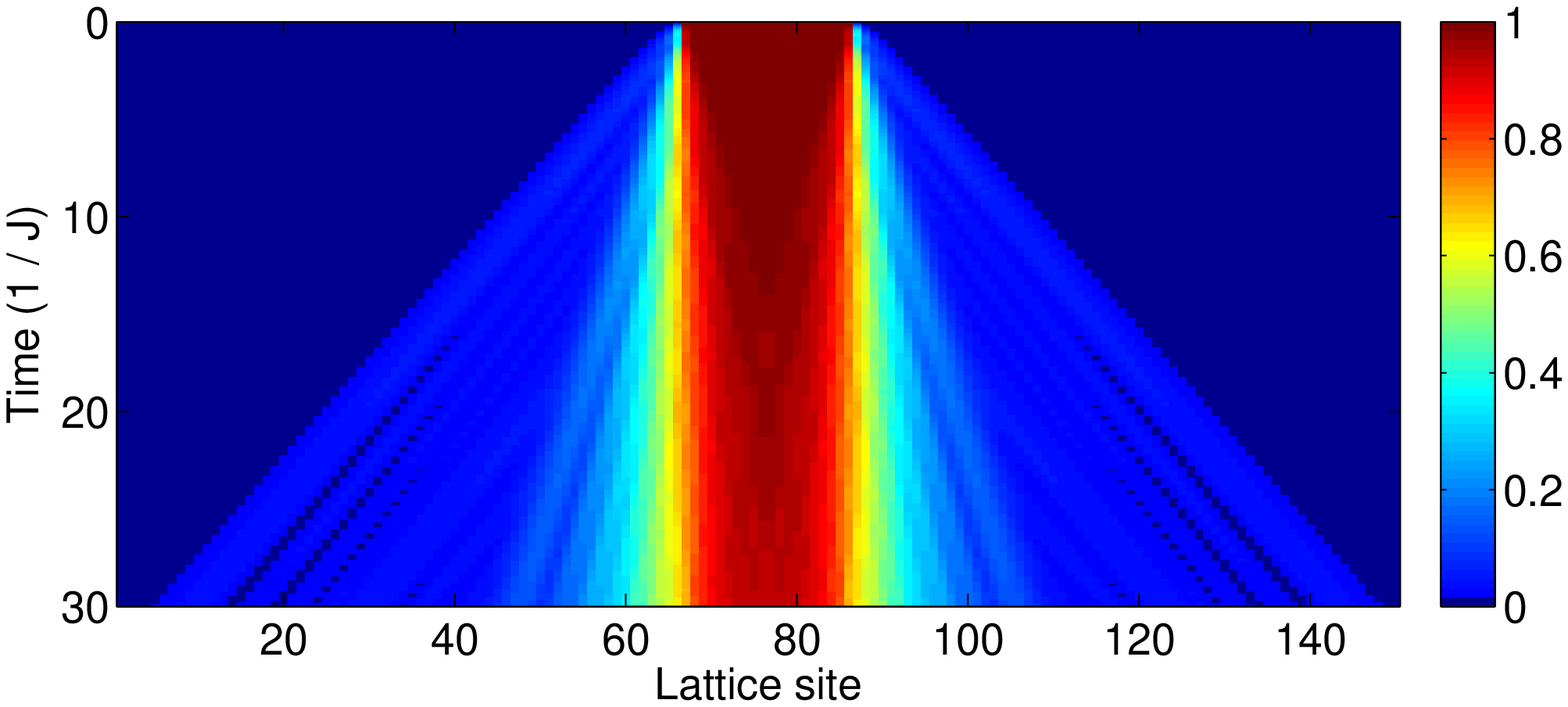}
        \includegraphics[width=0.45\textwidth]{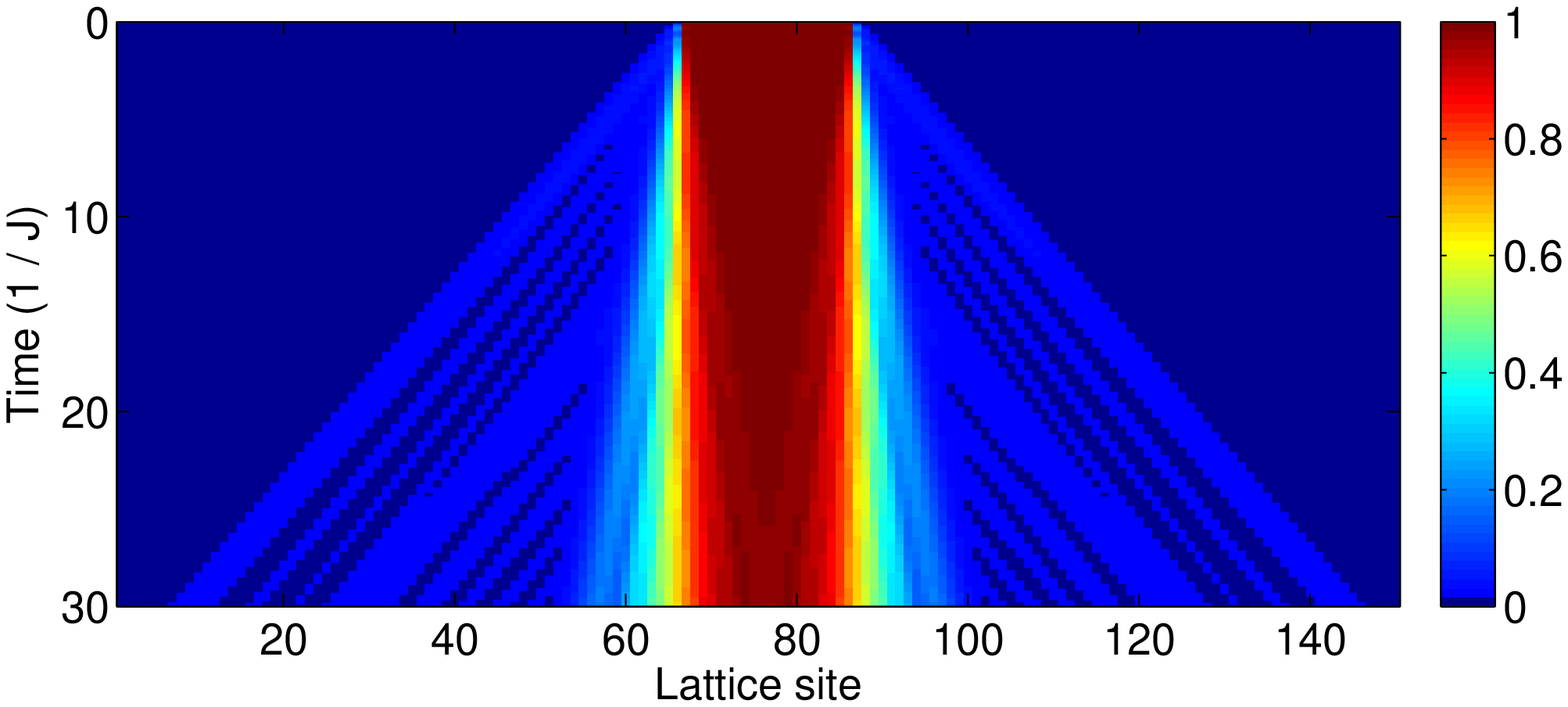}
	\begin{center}
          \caption{The square root of up density profiles $\sqrt{n_{i\, \uparrow} (t)}$. $|U| = 0,
            0.1, 1.0, 5.0, 10.0$ (top left to bottom right)}
          \label{fig:sqrt_n}
	\end{center}
       \end{figure}

\begin{figure}[ha]
        \includegraphics[width=0.45\textwidth]{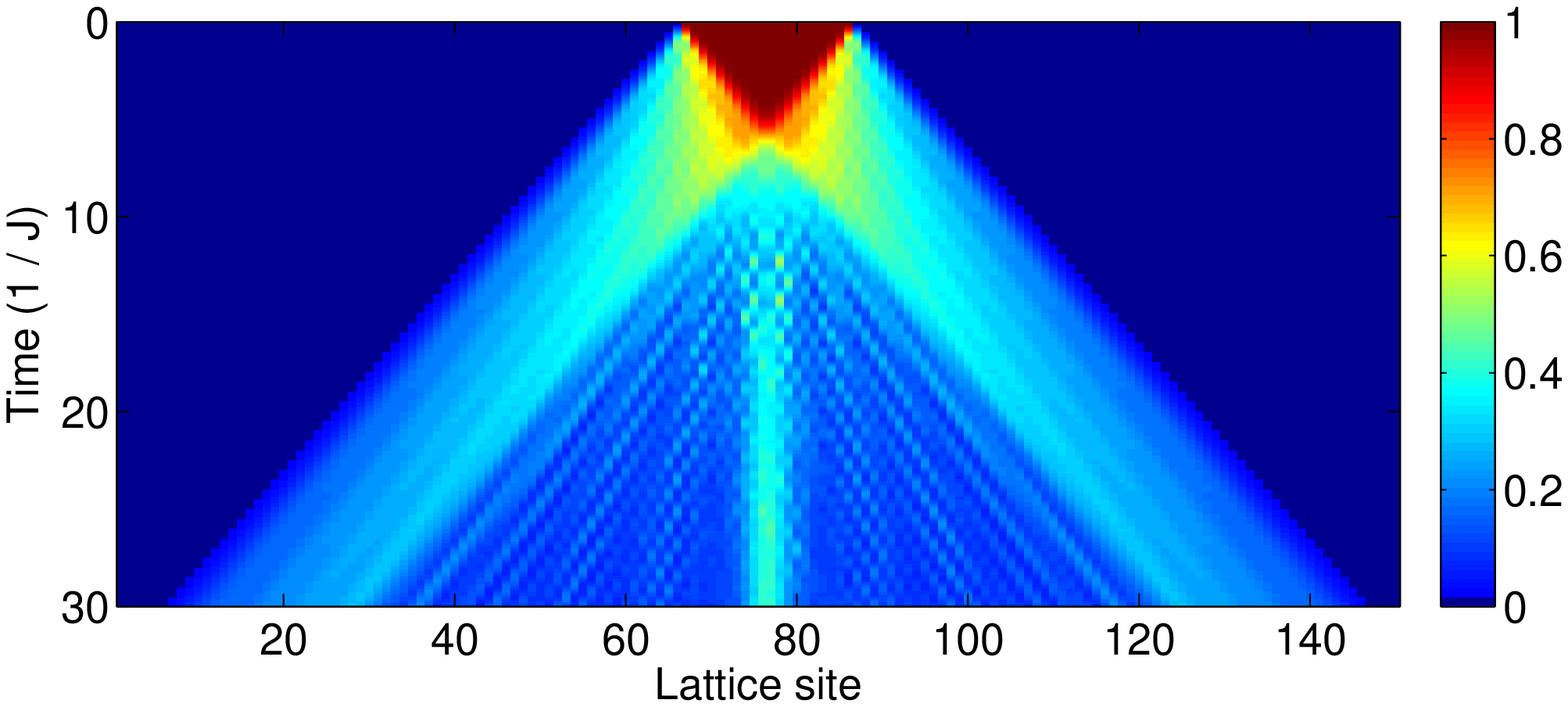}
	\includegraphics[width=0.45\textwidth]{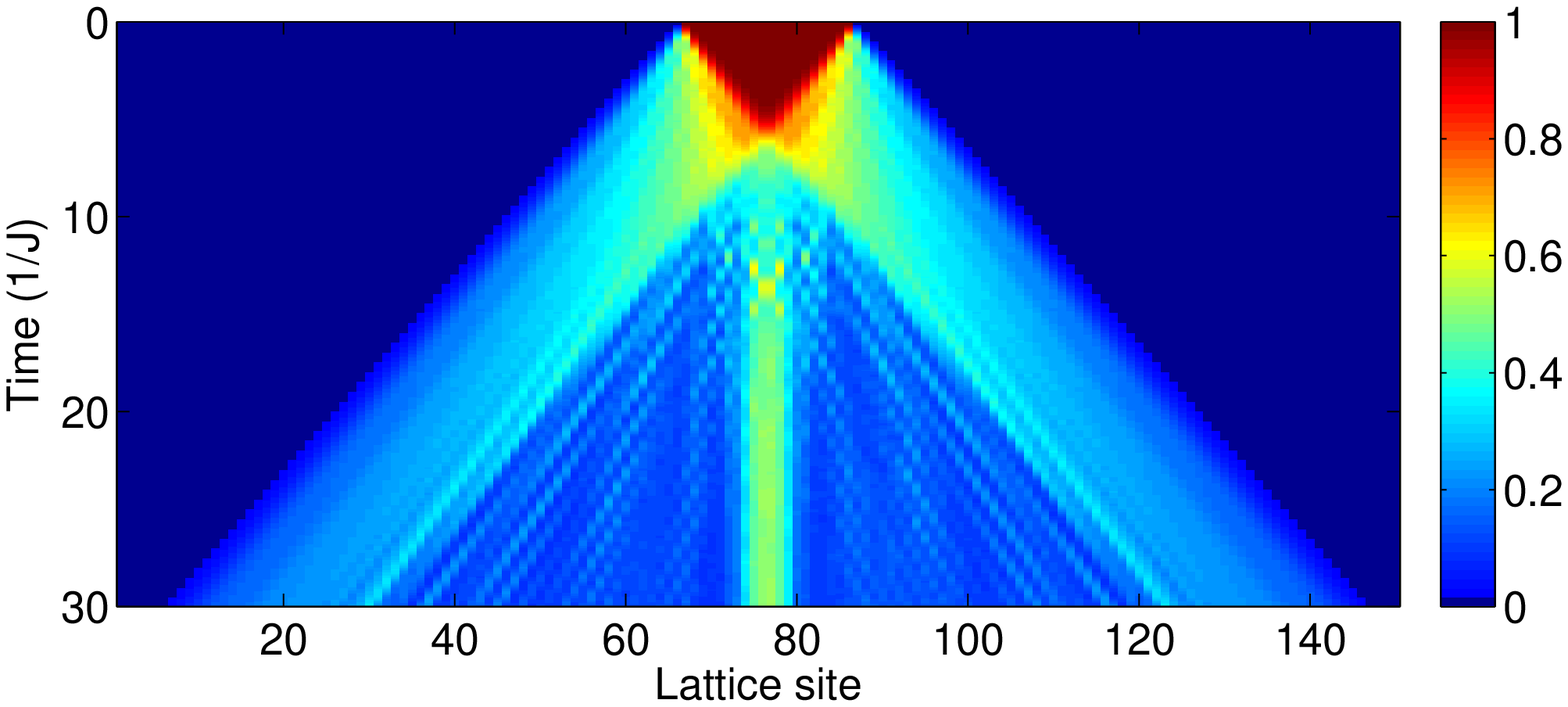}
        \includegraphics[width=0.45\textwidth]{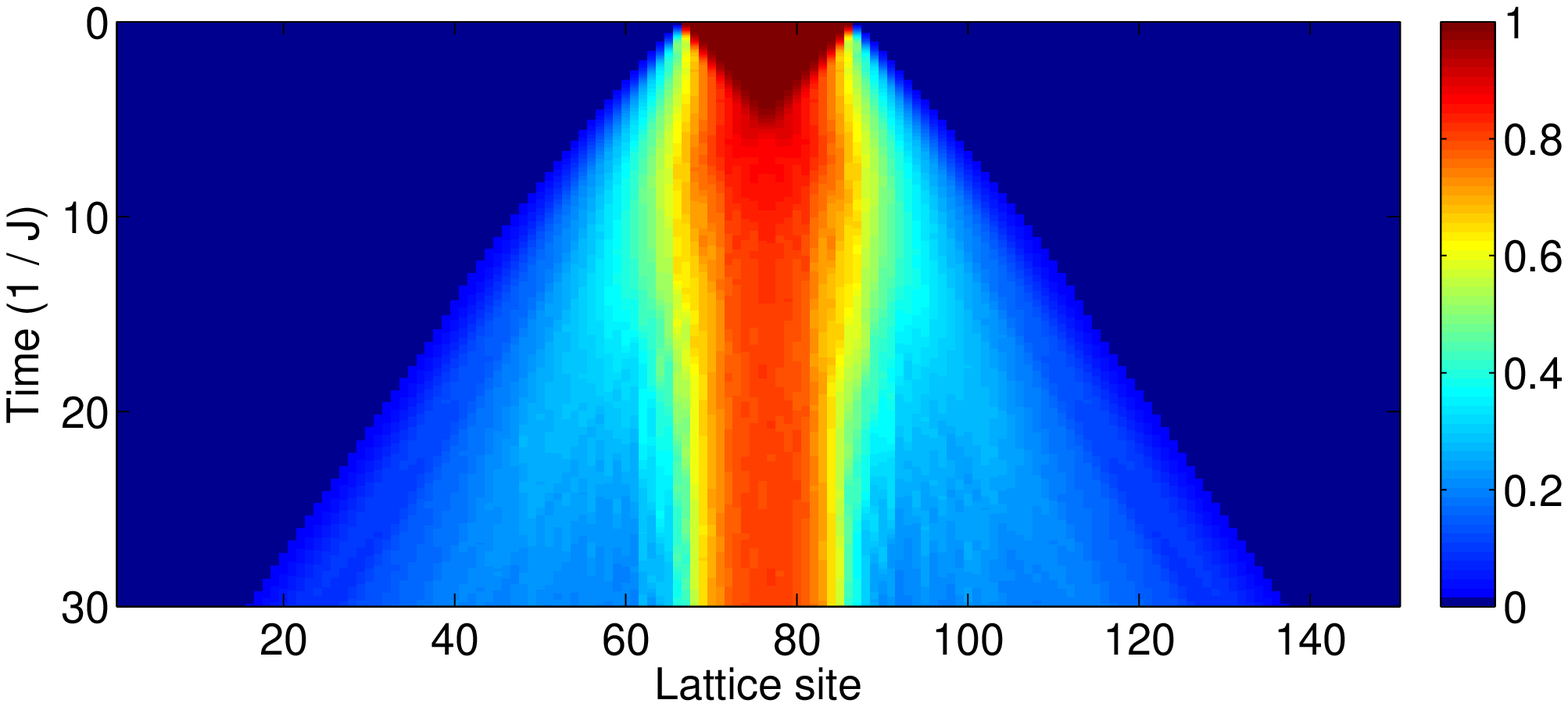}
        \includegraphics[width=0.45\textwidth]{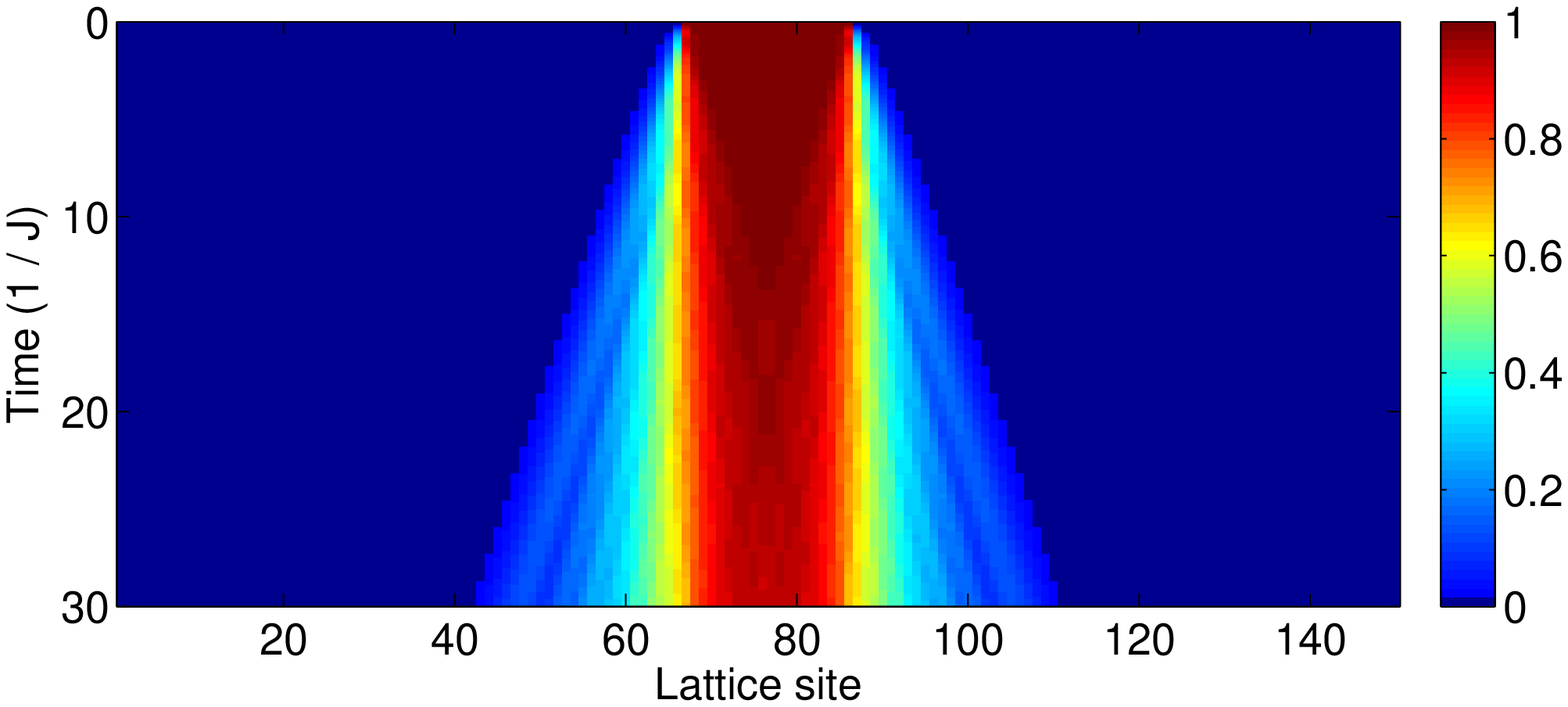}
        \includegraphics[width=0.45\textwidth]{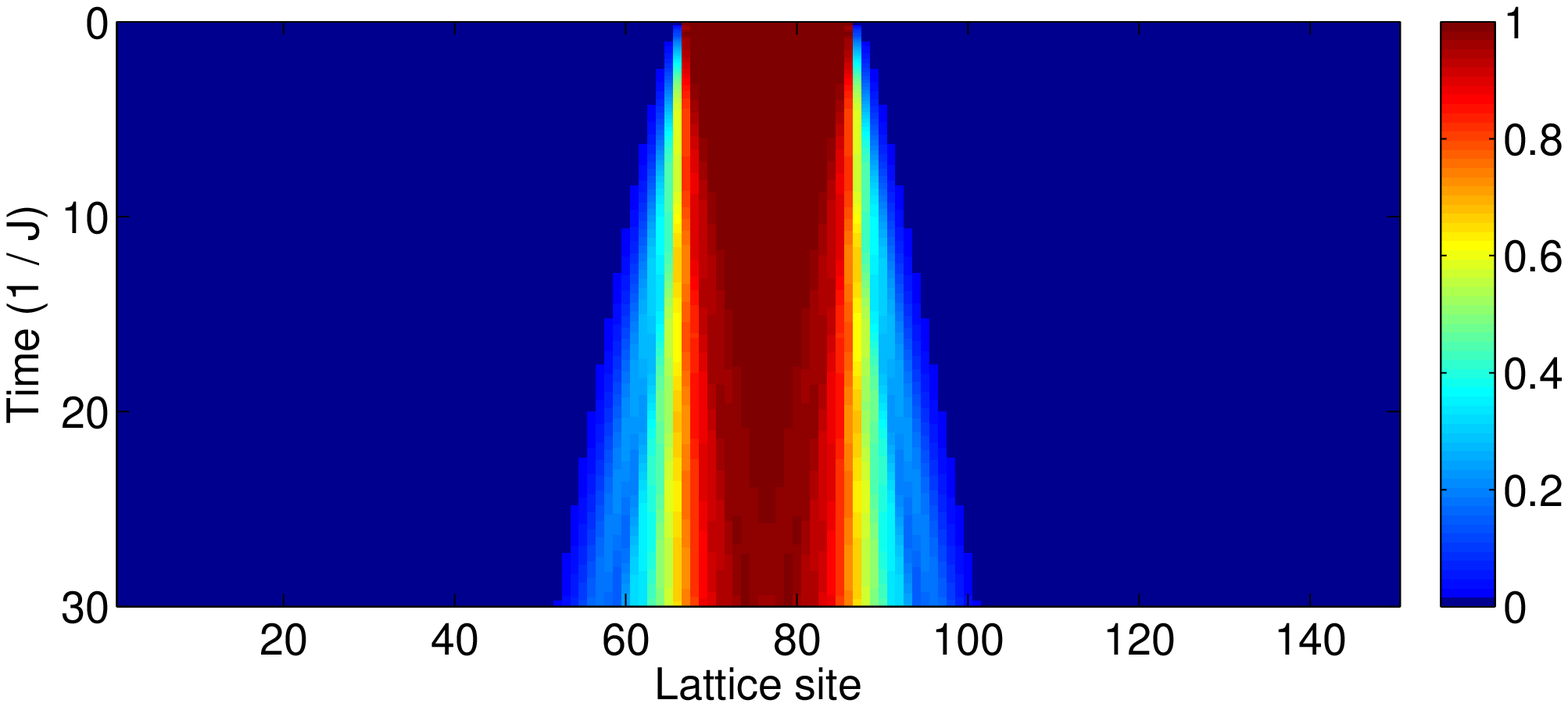}
	\begin{center}
          \caption{The square root of doublon density profiles. $|U| =
            0, 0.1, 1.0, 5.0, 10.0$ (left to right, top to bottom)}
          \label{fig:sqrt_doubl}
       \end{center}
      \end{figure}

\begin{figure}[!h]
         \includegraphics[width=0.45\textwidth]{U_10up.eps}
         \includegraphics[width=0.45\textwidth]{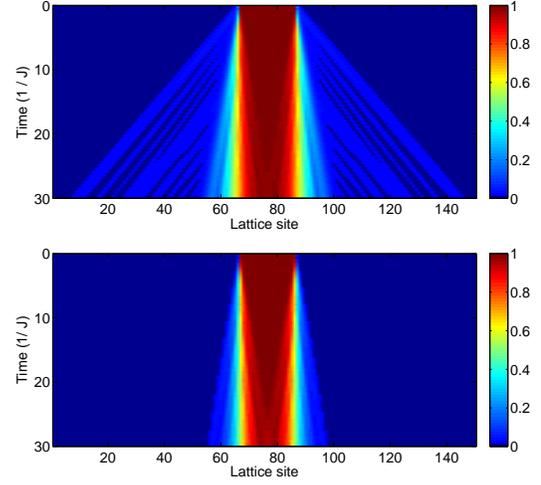}
         \begin{center}
           \caption{Comparison between the square root of the density
             profile $\sqrt{n_{i\, \uparrow} (t)}$ (left) with the
             density profile $n_{i\, \uparrow} (t)$(right) for $|U| =
             10.0$.The unpaired particle wavefronts expanding with
             speed $2J$ are clearly visible in the former.}
          \label{fig:comp_n}
       \end{center}
      \end{figure}

\begin{figure}[!h]
         \includegraphics[width=0.45\textwidth]{U_10updown.eps}
         \includegraphics[width=0.45\textwidth]{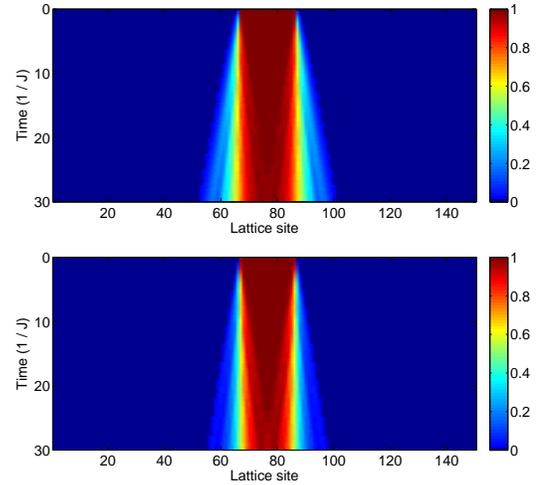}
       	\begin{center}
          \caption{Comparison between the square root of the doublons
            density profile $\sqrt{n_{i\, \uparrow \downarrow} (t)}$
            (left) with the density profile $n_{i\, \uparrow
              \downarrow} (t)$(right) for $|U| = 10.0$. In the left
            panel ( $\sqrt{n_{i\, \uparrow\downarrow} (t)}$)
            low-density features are enhanced.}
          \label{fig:comp_doubl}
       \end{center}
      \end{figure}

We show in Figs. \ref{fig:sqrt_n} and \ref{fig:sqrt_doubl} the time
evolution of the (square-root) density for unpaired particles
$\sqrt{n_{i\, \uparrow} (t)}$ and doublons $\sqrt{n_{i\, \uparrow
    \downarrow} (t)}$ for different values of the interaction.  While
the outermost wavefront velocity ($2J$), being related to the
expansion of the unpaired particles, in Fig. \ref{fig:sqrt_n} is
interaction-independent, the doublon wavefronts in
Fig. \ref{fig:sqrt_doubl} have an interaction-dependent velocity
($4J^2/U$). In order to give grounds for our choice to plot the square
root of unpaired particles and doublons in the article, in
Figs. \ref{fig:comp_n} and \ref{fig:comp_doubl} we compare the square root
densities and the densities for $|U|=10.0$. It is noted that we have studied 
the time evolution up to time t = 30 1/J. At this time the expanding cloud has
not reached the edges of our finite system. However, the finite size causes
low-momentum cutoff in the expansion dynamics. Therefore, we have 
experimented with systems of different size and we have seen that the 
lattice size does not affect the expansion dynamics significantly.

\subsection{Doublon and unpaired particle oscillations}
\label{sec:freq_osc}
\begin{figure}[H]
      \includegraphics[width=0.45\textwidth]{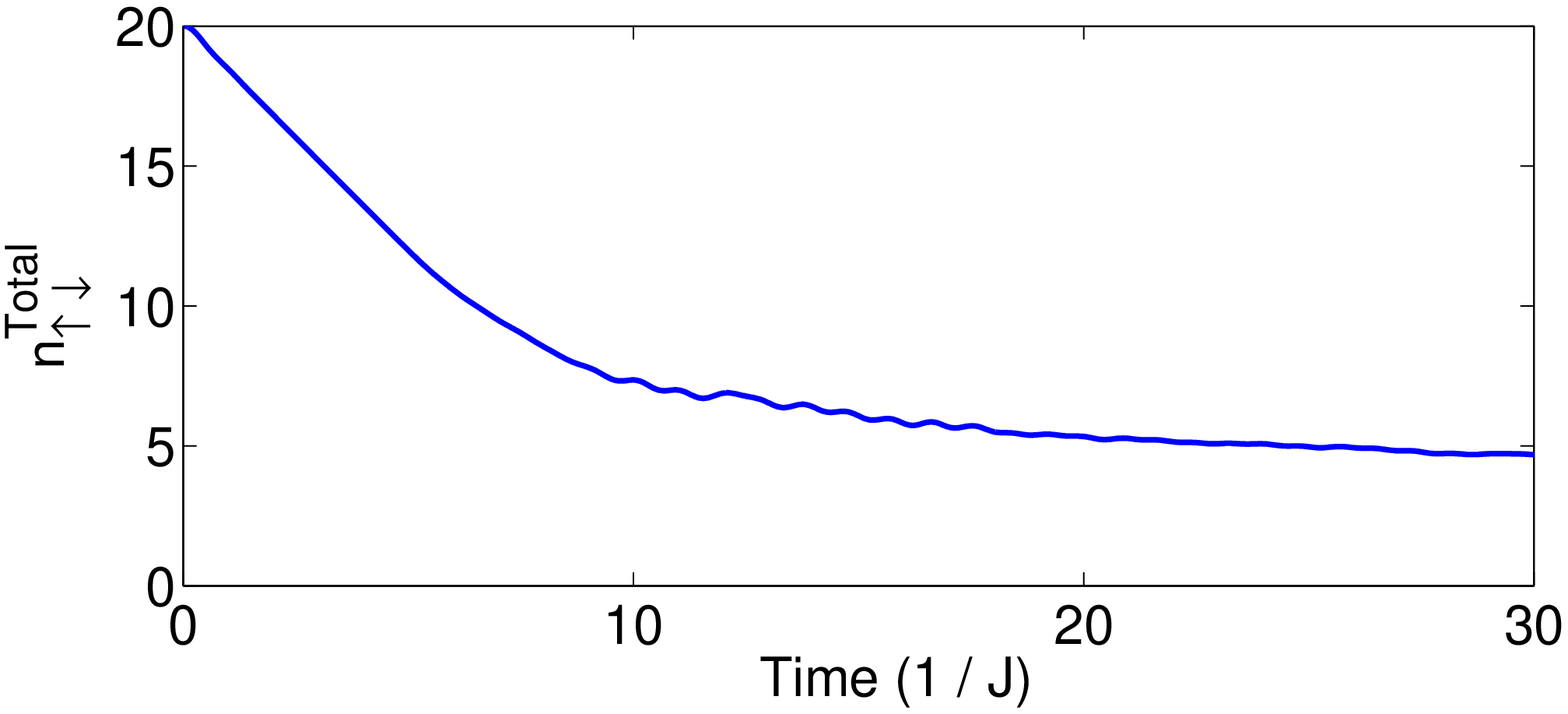}	
      \includegraphics[width=0.45\textwidth]{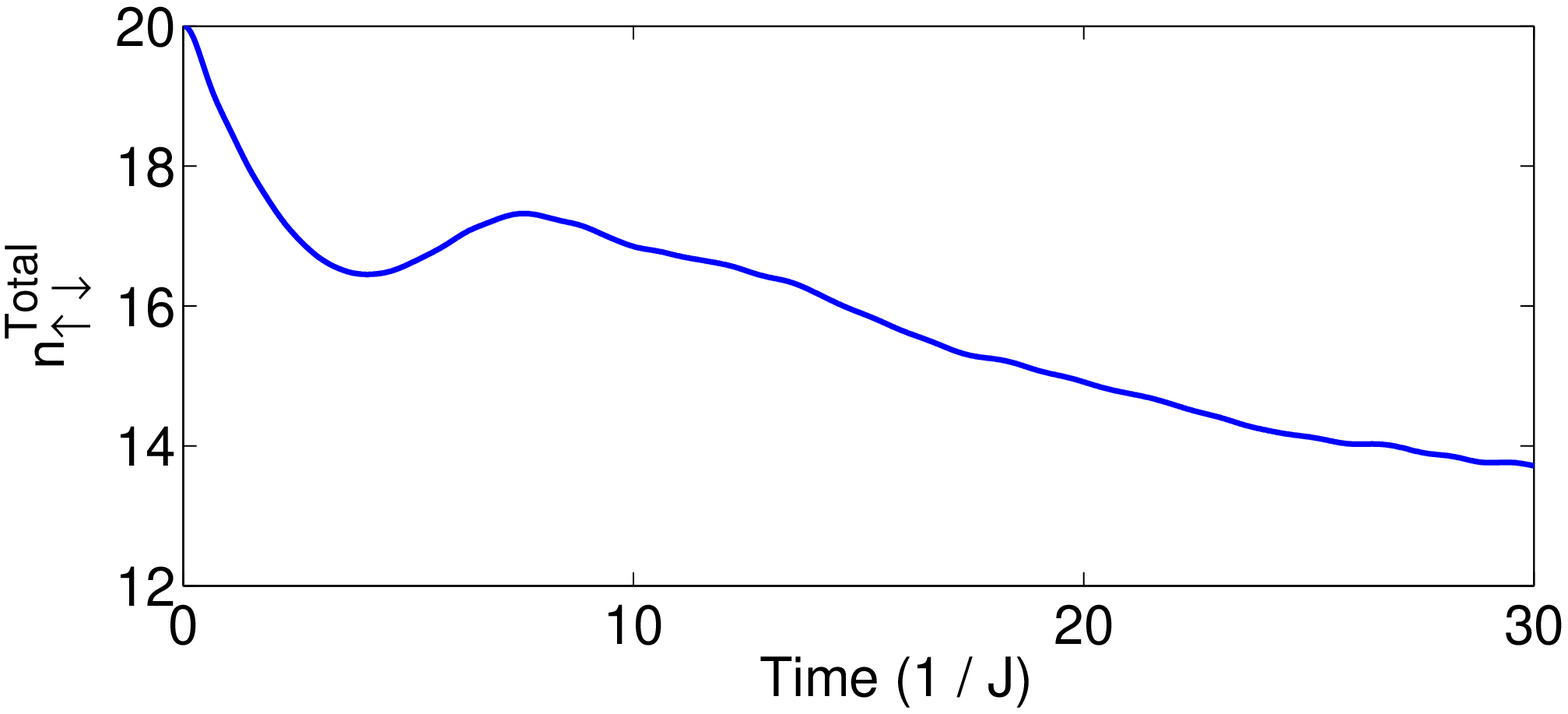}
       \includegraphics[width=0.45\textwidth]{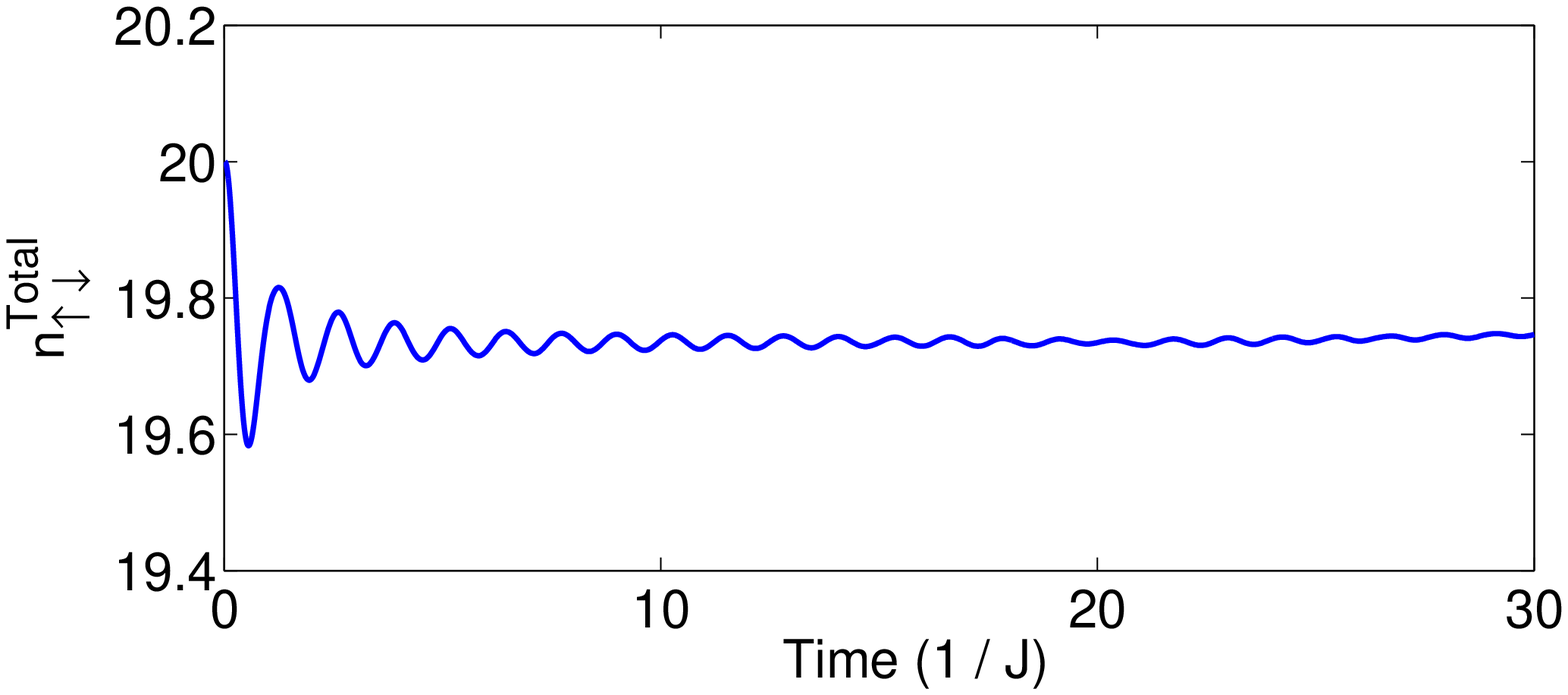}	
      \includegraphics[width=0.45\textwidth]{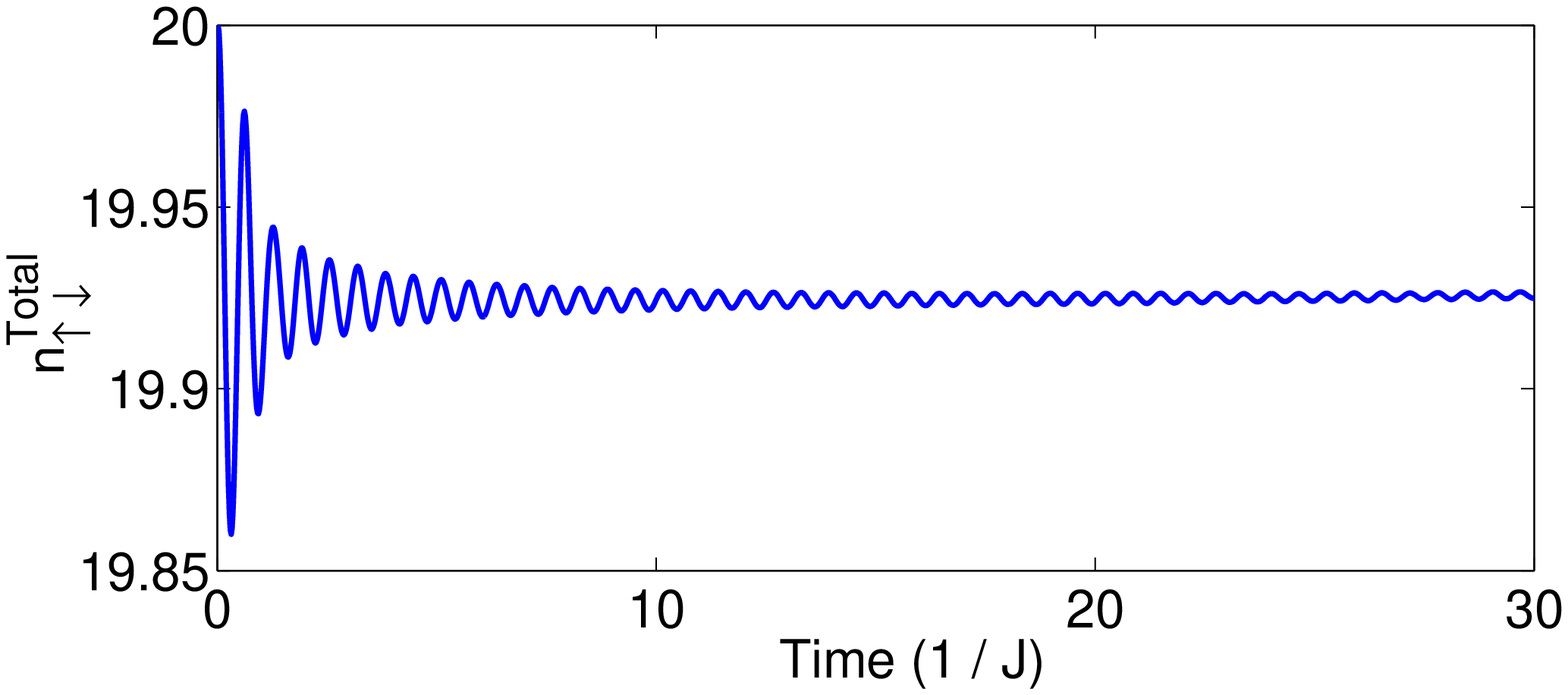}
      \begin{center}
	\caption{The time dependence of $n^{Total}_{i, \uparrow
            \downarrow} (t)$ for $|U|=0.1, 1.0, 5.0, 10.0$ (top left to
           bottom right)}
       \label{fig:Dt}
      \end{center}
    \end{figure}

\begin{figure}[H]
      \includegraphics[width=0.45\textwidth]{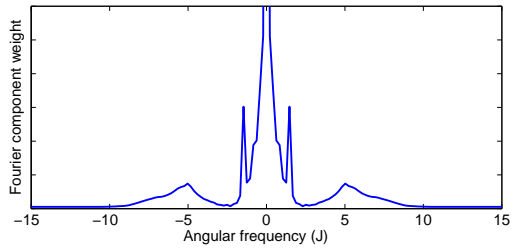}
      \begin{center}
	\caption{The Fourier transform of $n_{E_L+O_L,\uparrow \downarrow}
          (t)$ for U=5.0.}
       \label{fig:D_FT}
      \end{center}
    \end{figure}

\begin{figure}[!h]
      \includegraphics[width=0.45\textwidth]{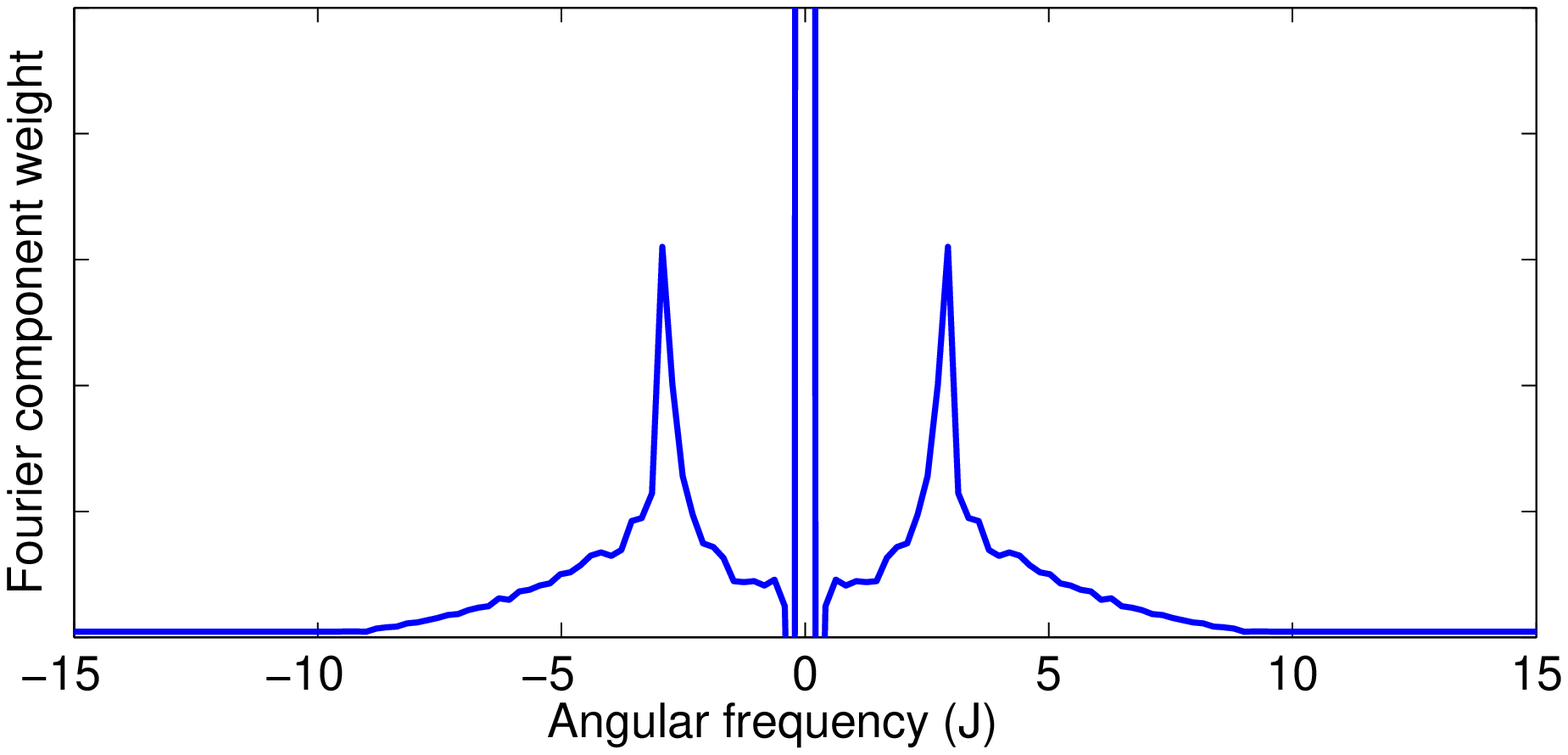}	
      \includegraphics[width=0.45\textwidth]{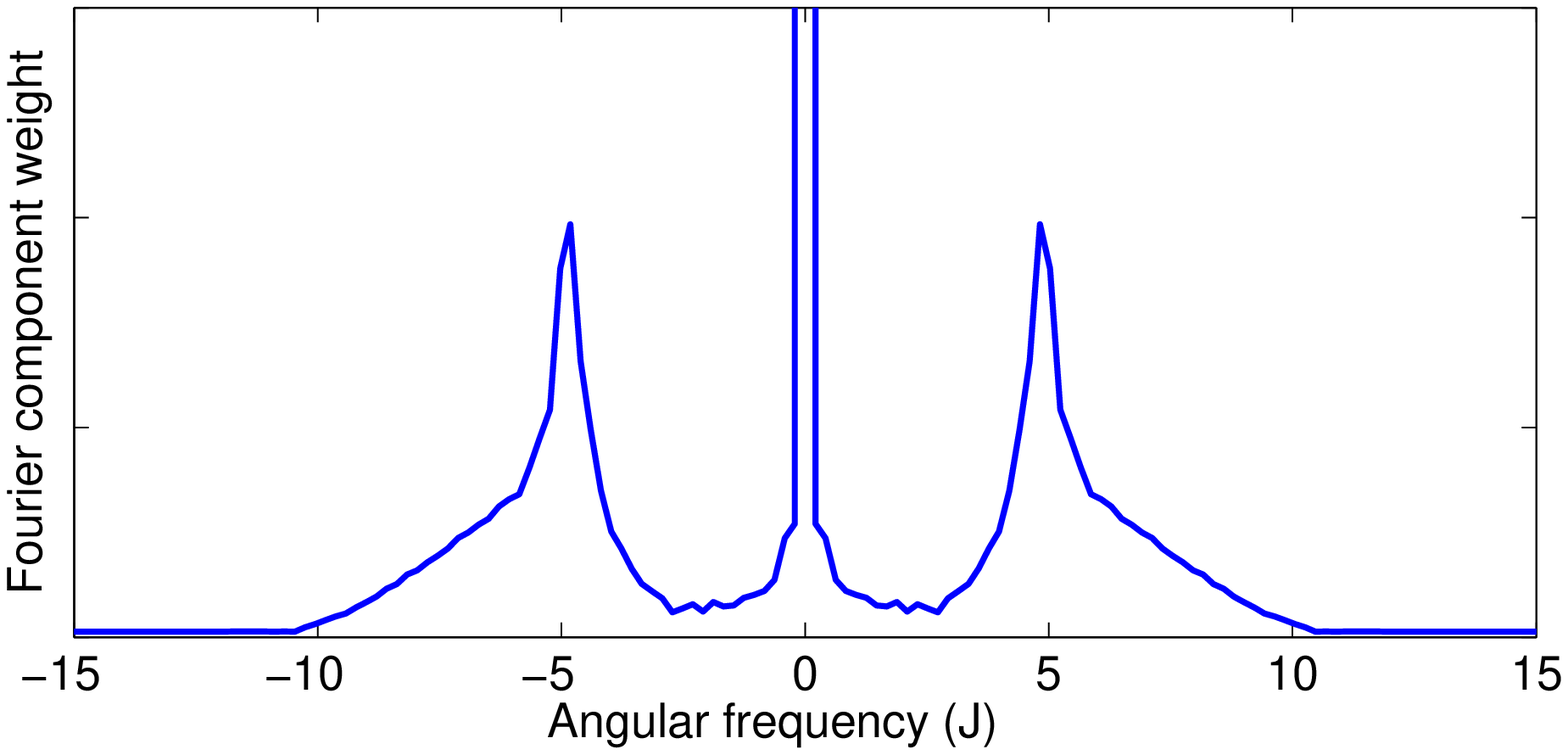}
      \begin{center}
	\caption{Fourier transform of $n^{Total}_{\uparrow \downarrow}
          (t)$ for U=3.0 (left) and U=5.0 (right)}
       \label{fig:Dt_FT}
      \end{center}
\end{figure}

In the main article, we compare the oscillation of the unpaired population
in the Hubbard dimer $ \tilde{n}_{E_L+O_L, \uparrow}^{un} (t)$ to its
counterpart obtained with the TEBD algorithm $n_{E_L+O_L, \uparrow}^{un}
(t)$. In Fig. \ref{fig:Dt} the number of doublons in the whole lattice
$n^{Total}_{i, \uparrow \downarrow} (t)$ as a function of time for
various interactions are shown, exhibiting increasing oscillation
frequency with increasing $|U|$. In Figs.  \ref{fig:D_FT} and
\ref{fig:Dt_FT} we show the Fourier transform (FT) of $n_{E_L+O_L,
  \uparrow \downarrow}(t)$ and of $n^{Total}_{i, \uparrow \downarrow}
(t)$ for $|U|=5$.  Owing to particle-number conservation $n_{E_L+O_L,
  \uparrow}^{un} (t)$ shows the same time oscillatory dependence as
$n_{E_L+O_L, \uparrow \downarrow}(t)$, although naturally in the opposite
phase. The presence of extra peaks present in the FT of $n_{E_L+O_L,
  \uparrow \downarrow}(t)$ with respect to the FT of $n_{E_L+O_L,
  \uparrow}^{un} (t)$ will be explained later.


 \begin{figure}[!h]
      \includegraphics[width=0.45\textwidth]{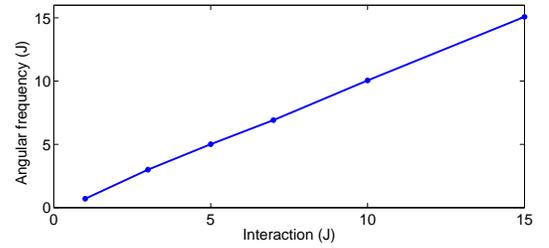}	
      \begin{center}
	\caption{The position of the peaks in the Fourier transform of
          $n_{E_L\,\uparrow \downarrow}(t)$ as a function of the
          interaction strength.}
    \label{fig:pks}
      \end{center}
 \end{figure}

In the numerics the FT $n_{E_L+O_L, \uparrow \downarrow}^{un} (t)$ shows
extra peaks with respect to the dimer dynamics which are associated to
the hopping between the dimer and the rest of the chain. The peak at
$\omega \simeq U$, present both for $n^{Total}_{\uparrow \downarrow}
(t)$ and $n_{E_L+O_L,\uparrow \downarrow} (t)$, correspond to the doublon
dissociation frequency, as it will be later shown.

\section{The Hubbard dimer model}
\label{sec:dim_mod}

The eigenstates of the two-site Hubbard Hamiltonian (so called Hubbard
dimer)

\begin{eqnarray}
      \label{eq:HubbHam}
        H=H_J+H_{int} \nonumber \\
       H_J=-J \sum_\sigma c^\dagger_{1,\sigma} c_{2\,\sigma} + h.c. \nonumber  \\
       H_{int}= U\sum_{i=1,2} n_{i\,\uparrow}n_{i\,\downarrow}
\end{eqnarray}
are given by (see e.g. \cite{Trotzky:2008p2201})
\begin{eqnarray}
  \label{eq:symmEig}
  &&|T\rangle  =\frac{1}{\sqrt{2}}\left( | \uparrow, \downarrow\rangle + | \downarrow, \uparrow\rangle \right) 
  \nonumber \\
  &&|D_-\rangle=\frac{1}{\sqrt{2}}\left( | \uparrow \downarrow,0\rangle - | 0, \uparrow \downarrow\rangle \right)
  \nonumber \\
  &&|v_{\pm}\rangle=\frac{1}{\sqrt{1+\alpha_{\pm}^2}}\left(|S\rangle+\alpha_{\pm}|D_+\rangle\right),
\end{eqnarray}
where
 \begin{eqnarray}
   \label{eq:alpha,S,Dp}
      \alpha_{\pm}=-\frac{U}{4J}\left[1 \pm \sqrt{1+\frac{16 J^2}{U^2}}\right],\\
     |S\rangle  =\frac{1}{\sqrt{2}}\left( | \uparrow, \downarrow\rangle - | \downarrow, \uparrow\rangle \right) \\
     |D_+\rangle= \frac{1}{\sqrt{2}}\left( | \uparrow \downarrow,0\rangle + | 0, \uparrow \downarrow\rangle \right)
  \end{eqnarray}
and the corresponding eigenvalues have the following form
  \begin{equation}
      \label{eq:spectr}
      \begin{array}{ll}
      \lambda_-=U/2\left[1 - \sqrt{1+\frac{16 J^2}{U^2}} \right] (<0) \qquad &\Leftrightarrow
      \qquad |v_-\rangle \\
      \lambda_0=0    \qquad &\Leftrightarrow \qquad  |T\rangle  \\
      \lambda_U=U    \qquad &\Leftrightarrow \qquad  |D_-\rangle \\
      \lambda_+=U/2\left[1 + \sqrt{1+\frac{16 J^2}{U^2}} \right] (>U) \qquad &\Leftrightarrow
      \qquad |v_+\rangle.
     \end{array}
  \end{equation}

\subsection{Hubbard dimer dynamics}

We can apply the approach described in the previous section to the
analysis of the edge sites of the cloud. As described in the main
article, the edges of the cloud can be described as Hubbard dimers.
If we focus on the system $E_L+O_L$, the initial state is given by
 \begin{equation}
      \label{eq:initial}
      |\phi (t=0)\rangle  = | 0, \uparrow \downarrow\rangle=
      \frac{1}{\sqrt{2}}(\gamma_+ |v_+\rangle - \gamma_- |v_-\rangle - |D_-\rangle)
  \end{equation} 
where we have denoted
 \begin{equation}
      \label{eq:gamma}
      \gamma_{\pm} = \frac{1}{(\alpha_{+} - \alpha_{-})} \frac{1}{\sqrt{1+\alpha_{\pm}^2}}.
 \end{equation}
The number of doublons in the initially empty site $E_L$ is given by
\begin{eqnarray}
  \label{eq:doub_EL}
  n_{E_L+O_L,\uparrow \downarrow} (t) =& 
       \langle 0, \uparrow \downarrow|
           e^{\frac{i\hat{H}t}{\hbar}} \,  \hat{n}^{L}_{\uparrow \downarrow} \, e^{-\frac{i\hat{H}t}{\hbar}} 
       |0, \uparrow \downarrow \rangle  \nonumber \\
      =& \frac{1}{4} + \frac{\alpha^2_+ + \alpha^2_-}{4(\alpha_+ - \alpha_-)^2}  
         + \frac{1}{2(\alpha_+ - \alpha_-)^2} \nonumber \\
	& - \left[ \frac{\alpha_+ \alpha_-+1}{2(\alpha_+ - \alpha_-)^2 }\right] \cos\left[(\lambda_+ - \lambda_-)t\right] \nonumber \\
	& -\frac{\alpha_+}{2\left(\alpha_+ - \alpha_-\right)}\cos\left[(\lambda_U - \lambda_+)t\right] \nonumber \\
	& + \frac{\alpha_-}{2\left(\alpha_+ - \alpha_-\right) }\cos\left[(\lambda_U - \lambda_-)t\right]. 
\end{eqnarray}
Straightforwardly, Eq.~(\ref{eq:doub_EL}) allows to evaluate the number
of unpaired particles in the same site
\begin{eqnarray}
  \label{eq:n_EL}
  n_{E_L+O_L,\uparrow} (t) =&  \frac{1}{4} + \frac{\alpha^2_+ + \alpha^2_-}{4(\alpha_+ - \alpha_-)^2} + \frac{1}{2(\alpha_+ - \alpha_-)^2} \nonumber \\
	& - \left[\frac{\alpha_+ \alpha_-}{2(\alpha_+ - \alpha_-)^2 }+\frac{1}{2(\alpha_+ - \alpha_-)^2 }\right] \nonumber\\
        & * cos\left[(\lambda_+ - \lambda_-)t\right] \nonumber \\
	& - \frac{\alpha_+}{2\left(\alpha_+ - \alpha_-\right) }\cos\left[(\lambda_U - \lambda_+)t\right] \nonumber \\
	& + \frac{\alpha_-}{2\left(\alpha_+ - \alpha_-\right) }\cos\left[(\lambda_U - \lambda_-)t\right].
\end{eqnarray}

The extra peaks appearing in the FT of $n_{E_L+O_L,\uparrow\downarrow}
(t)$ (Fig. \ref{fig:D_FT}), with respect to the FT of
$n^{Total}_{\uparrow\downarrow} (t)$ (Fig. \ref{fig:Dt_FT} are
due to the fact that, on top of the doublon dissociation
dynamics, $n_{E_L+O_L,\uparrow\downarrow} (t)$ is affected by the
hopping of both unpaired particles and doublons in and out of the
dimer $E_L/O_L$. These contributions, since a sum over the whole cloud
is carried out, add up to zero for $n^{Total}_{\uparrow\downarrow} (t)$.

\section{Expansion velocities for particles and doublons}
\label{sec:exp_vel}

The expansion velocity of the unpaired particles can be easily
obtained considering the dispersion relation of a particle hopping
through on an empty lattice
\begin{equation}
  \label{eq:un_disp}
  \epsilon_k=-2J\cos(k)
\end{equation}
leading to the group velocity
\begin{equation}
  \label{eq:v_un}
  v_k=\frac{\partial \epsilon_k}{\partial k}=2J\sin(k).
\end{equation}
Since in our system we consider particles generated at a specific site
($E_L$), all momentum states contribute to the expression of the
initial state. From Eq. (\ref{eq:v_un}) we thus see that the maximum velocity is attained for 
$k=\pi/2$
\begin{equation}
  \label{eq:vm_un}
  v^{un}_{max}=2J\sin(k)\big|_{k=\frac{\pi}{2}}=2J
\end{equation}
which corresponds to the value obtained in the TEBD simulation.
Analogously it can be shown that, for our initial configuration, the
doublon hopping corresponds to the hopping of a particle with
$J_{eff}=\frac{4 J^2}{U}$
\cite{Giamarchi:2004p1570,HLEssler:2005p1368,Winkler:2006p60}
%%%%francesco FROM HERE %%%%%%%%,
corresponding to the velocity of propagation of a charge excitation in
the strongly attractive Hubbard model. In particular, if we consider
the case where no unpaired particles are present, we can map our
system to an isotropic Heisenberg chain by identifying $|\uparrow
\downarrow\rangle \Rightarrow |\uparrow\rangle$ and
$|\emptyset\rangle\Rightarrow |\downarrow\rangle$. To prove this,
we consider:
 \begin{enumerate}
 \item A staggered particle-hole transformation 
   \begin{align}
     \label{eq:stagger}
     &c^{(\dagger)}_{i\,\uparrow} \Rightarrow c^{(\dagger)}_{i\,\uparrow}\\
     &c^{\dagger}_{i\,\downarrow} \Rightarrow (-1)^i c_{i\,\downarrow}\\
     &c_{i\,\downarrow} \Rightarrow  (-1)^i c^{\dagger}_{i\,\downarrow}\\
   \end{align}
   which transforms our original system (filling
   $n=\sum_{i\,\sigma}n_{i\,\sigma}/L$, magnetization density $m=0$
   and interaction strength $U$) to a half-filled system with filling
   $n'=1$, magnetization density $m'=n-1$ and interaction strength
   $U'=-U$. Since $n'_{i\,\uparrow}=n_{i\,\uparrow\downarrow} \\
   n'_{i\,\downarrow}=1-n_{i\,\uparrow\downarrow}$;
  
  \item by strong coupling second order expansion
    \cite{HLEssler:2005p1368}, the system Hamiltonian is then mapped
    to an isotropic Heisenberg Hamiltonian
   \begin{equation}
     \label{eq:HH}
      H_{H}=-\frac{4 J^2}{U} \sum_i \overrightarrow{S}_i \overrightarrow{S}_{i+1}.
   \end{equation}
   With the mappings described above, since a doubly occupied site
   correspond to a spin up, and an empty site to a spin down, the
   band-insulator initial state $\left| \phi(0) \right\rangle$ is thus
   transformed to a spin chain with $O_R-O_L$ central spin-up sites
   surrounded by spin-down sites, and the expansion dynamics of the
   doublons corresponds to the propagation of a single spin-up
   excitation through a spin-down polarized system.  Its dispersion
   relation is given by
  \begin{equation}
    \label{eq:e_k}
    \epsilon_k^{spin}=\frac{4 J^2}{U} \cos(k)
  \end{equation}
  and hence, analogously to the derivation of the maximum velocity for the
  unpaired particles, to 
  \begin{equation}
    \label{eq:vdoubl}
    v_{MAX}^{doublon}=v_{MAX}^{spin}=\left| \frac{4 J^2}{U}\right|.
  \end{equation}
 \end{enumerate}
%%%%%francesco TO HERE%%%%%%%%

\section{Definition of the core}
\label{sec:Dimer}

We have in the high interaction limit ballistic particles emitted from
the diffusive core, the two types of ballistic particles being
unpaired particles and doublons. These two expand at constant speeds
($2J$ and $\frac{4J^2}{U^2}$) in wavefronts whose densities are so low
that spin-spin scattering (i.e. the Hubbard Dimer dynamics) is
negligible.  In contrast, the core has high density, the Dimer
dynamics occur, and thus the core is diffusive. In the case of
interactions lower than $|U| < 3$ also the densities of the unpaired
and doublon wavefronts become high enough for the Dimer Dynamics,
which in the language of the classical non-linear diffusion model in
\cite{Schneider:2010p1468} makes the two types of wavefronts
diffusive. This diffusive behaviour is seen in our results as the
absence of ballistic wavefronts in density profiles (see e.g. the $|U|
= 1.0$ case in Fig.~\ref{fig:sqrt_n} above). Finally, for the
very low interactions $U \leq 0.5$ the problem simplifies to ballistic
expansion of non-interacting particles.

In order to compare our results to Fig. 5 of
\cite{Schneider:2010p1468} we arrive at the problem of defining the
core. In the high interaction limit this is perhaps easier as the
ballistic particles clearly separate from the central diffusive
region, the core. In the middle interaction region (see again the
$U=1.0$ case in Figs.~\ref{fig:sqrt_n} and \ref{fig:sqrt_doubl} below
) we see that only one outermost wavefront
expands ballistically and therefore if we consider the diffusive
region of the cloud to be the core then the core would be everything
else but the outermost wavefront. This definition however is
problematic in the low interaction limit (U $\leq$ 0.5).  In this
limit everything can be described as expansion of non-interacting
ballistic particles. So the definition of the core as the diffusive
part of the cloud does not work any more as then we would not have a
core at all.

However, experimentally it might be easier to define a high density region to correspond 
to the core, and fit the half-power half-maximum of a Gaussian to give the core width
and subsequently the core expansion speed. In effect, then, in the low interaction 
regime the core is defined as the whole cloud, and in the middle and high interaction 
regimes the core is defined as everything else except the ballistic expansion fronts. 
This is what we have done in the main article. 

\begin{figure}
\includegraphics[width=0.45\textwidth]{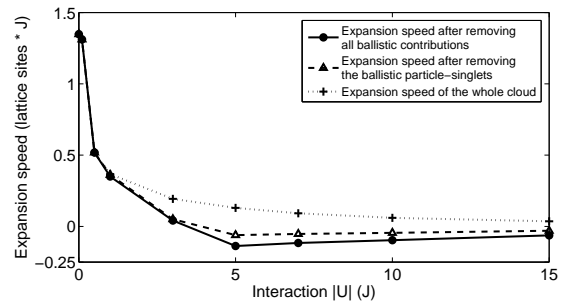} 
\caption {The core expansion velocities obtained by removing ballistic contributions (see text).}
\label{Fig:ExpansionSpeeds} 
\end{figure}

A variation of the above way of removing ballistic contributions would
be to remove only the ballistic \textit{particle} wavefronts, as the
ballistic hole wavefronts may in effect be included in the
experimental high density core region.  To elaborate on what the hole
wavefronts are, in the core state $|\uparrow \downarrow\rangle$ changes
into a superposition of $|\uparrow>$ and $|\downarrow\rangle$. This decreases
the total density, but in addition the $|\uparrow \rangle$ and $|\downarrow
\rangle$ states created are now part of a singlet state which expands
ballistically into the core. So, if we want to define the core to be
everything that is not ballistic, we need also to remove the density
contribution of the ballistic singlets in the core. Thus there are two
reasonable approaches in defining the core. First is that we remove
everything that is ballistic (particles and holes) and the second is
that we remove only the particles. The core expansion speeds obtained
in these two ways are plotted in \ref{Fig:ExpansionSpeeds}. Both
methods result in negative core expansion velocities as seen in the
experiment \cite{Schneider:2010p1468}.

\section{The time evolution of charge-charge correlations}
\label{sec:CCcorr}

Using TEBD we can calculate correlation functions during the expansion
of the cloud. An interesting correlation function in the case of the band insulator expansion
is the charge-charge correlator $C_{i j} (t)$, given by

\begin{equation} 
  \label{eq:ccCorr} 
  C_{i, j} (t) = <\Phi (t)| (\hat{n}_{i, \uparrow} + \hat{n}_{i, \downarrow})  (\hat{n}_{j, \uparrow} + \hat{n}_{j, \downarrow}) |\Phi (t)>.
\end{equation} 

However, $C_{i j} (t)$ is dominated by single particle density correlations. To see the effect of the non-trivial correlations,
we want to remove the density-density correlations from $C_{i, j} (t)$. If $i \neq j$ the density-density correlations are given by
$<\hat{n}_{i, \uparrow}> <\hat{n}_{ji, \uparrow}> + <\hat{n}_{i, \downarrow}> <\hat{n}_{j, \downarrow}> + <\hat{n}_{i, \uparrow}> <\hat{n}_{j, \downarrow}> + <\hat{n}_{i, \downarrow}> <\hat{n}_{j, \uparrow}>$.
When $i = j$, $<n_{i, \uparrow}n_{i, \uparrow}> = <c^{\dagger}_{i, \uparrow}c_{i, \uparrow}c^{\dagger}_{i, \uparrow}c_{i, \uparrow}> = <c^{\dagger}_{i, \uparrow}(1 - c^{\dagger}_{i, \uparrow}c_{i, \uparrow})c_{i, \uparrow}> = <c^{\dagger}_{i, \uparrow}c_{i, \uparrow}> = n_{i, \uparrow}$ and
therefore on the diagonal the we just remove the single particle and doublon densities. Therefore, the correlation function which displays
the non-trivial correlations is

\begin{eqnarray} 
  \label{eq:ccCorrD} 
D_{i, j} (t) = C_{i, j} (t) - (<\hat{n}_{i, \uparrow}> <\hat{n}_{j, \uparrow}> \nonumber \\ 
+ <\hat{n}_{i, \downarrow}> <\hat{n}_{j, \downarrow}> \nonumber \\
+ <\hat{n}_{i, \uparrow}> <\hat{n}_{j, \downarrow}> \nonumber \\ 
+ <\hat{n}_{i, \downarrow}> <\hat{n}_{j, \uparrow}>)\nonumber \\ 
(1 - \delta_{ij}) \nonumber  \\
- (<\hat{n}_{i, \uparrow}> + <\hat{n}_{i, \downarrow}> \nonumber \\
+ 2 <\hat{n}_{i, \uparrow \downarrow}>)\delta_{ij}. 
\end{eqnarray} 

\begin{figure}[!h]
\includegraphics[width=0.3\textwidth]{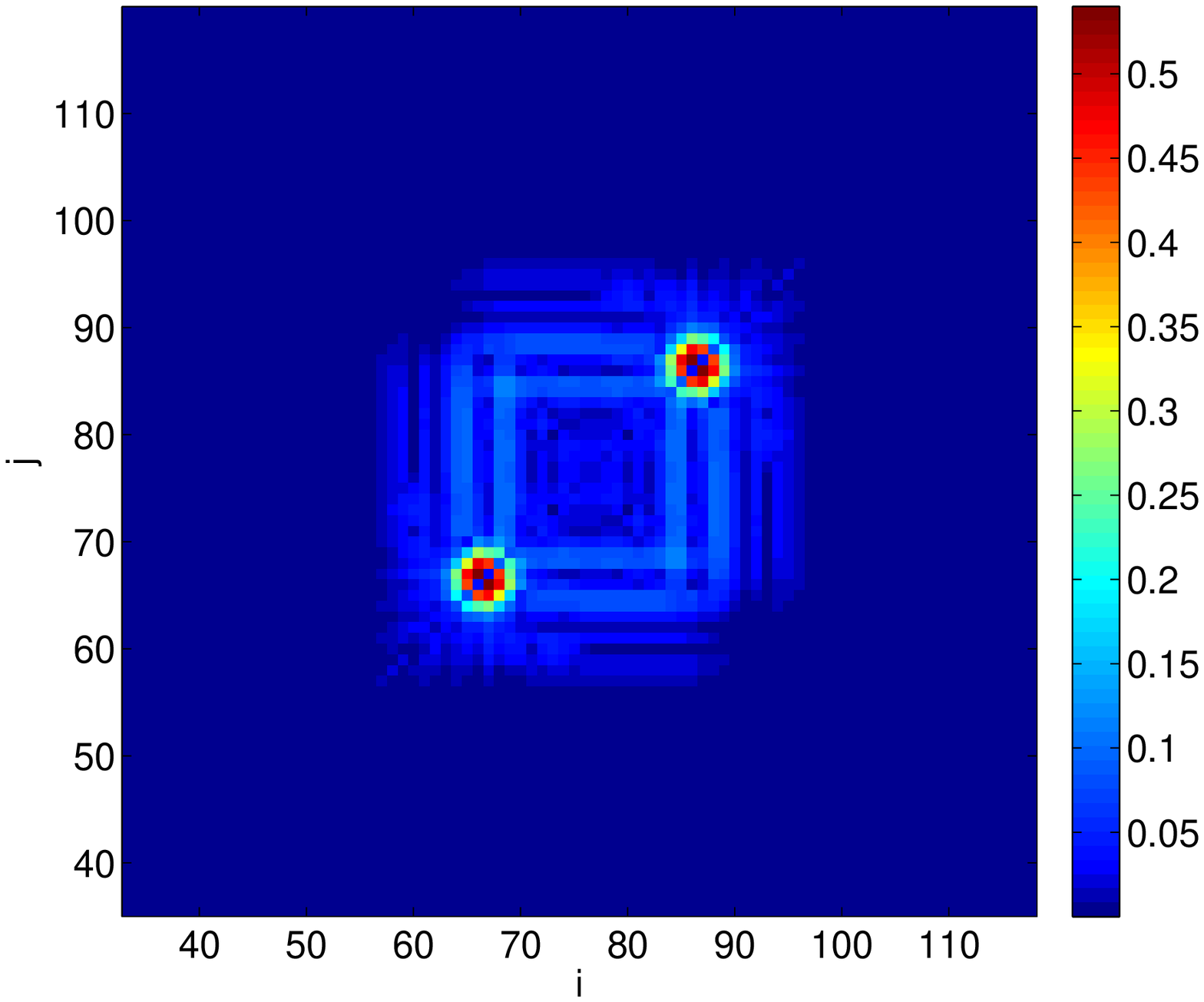} 
\includegraphics[width=0.3\textwidth]{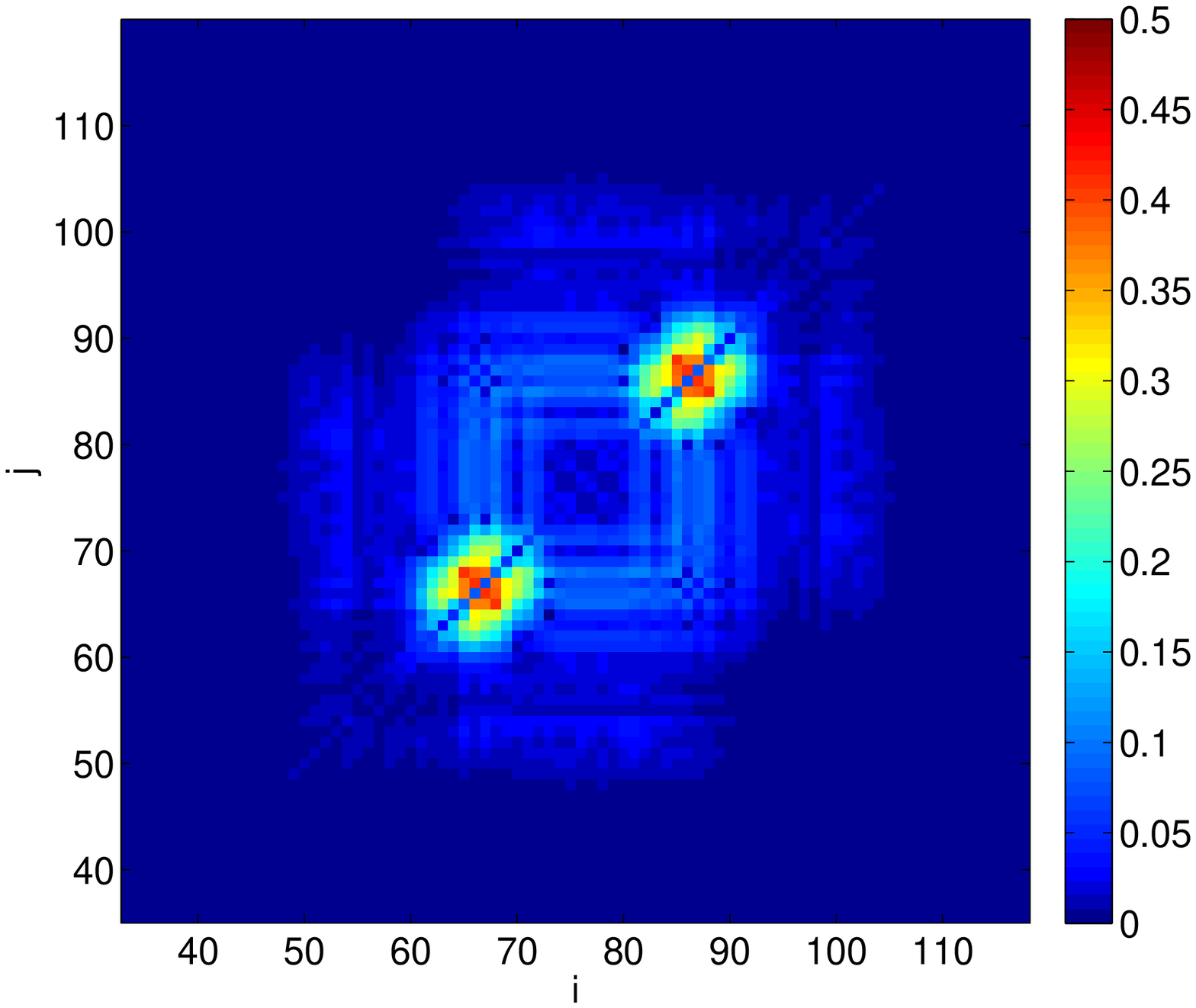}
\caption { (color online) Left:  $\sqrt{D_{i, j} (t)}$ 
for $|U| = 5.0$ at $t = 4 \frac{1}{J}$. Right: The same for $t = 8
\frac{1}{J}$.}
\label{fig:ccCorrA} 
\end{figure}

\begin{figure}[!h]
\includegraphics[width=0.3\textwidth]{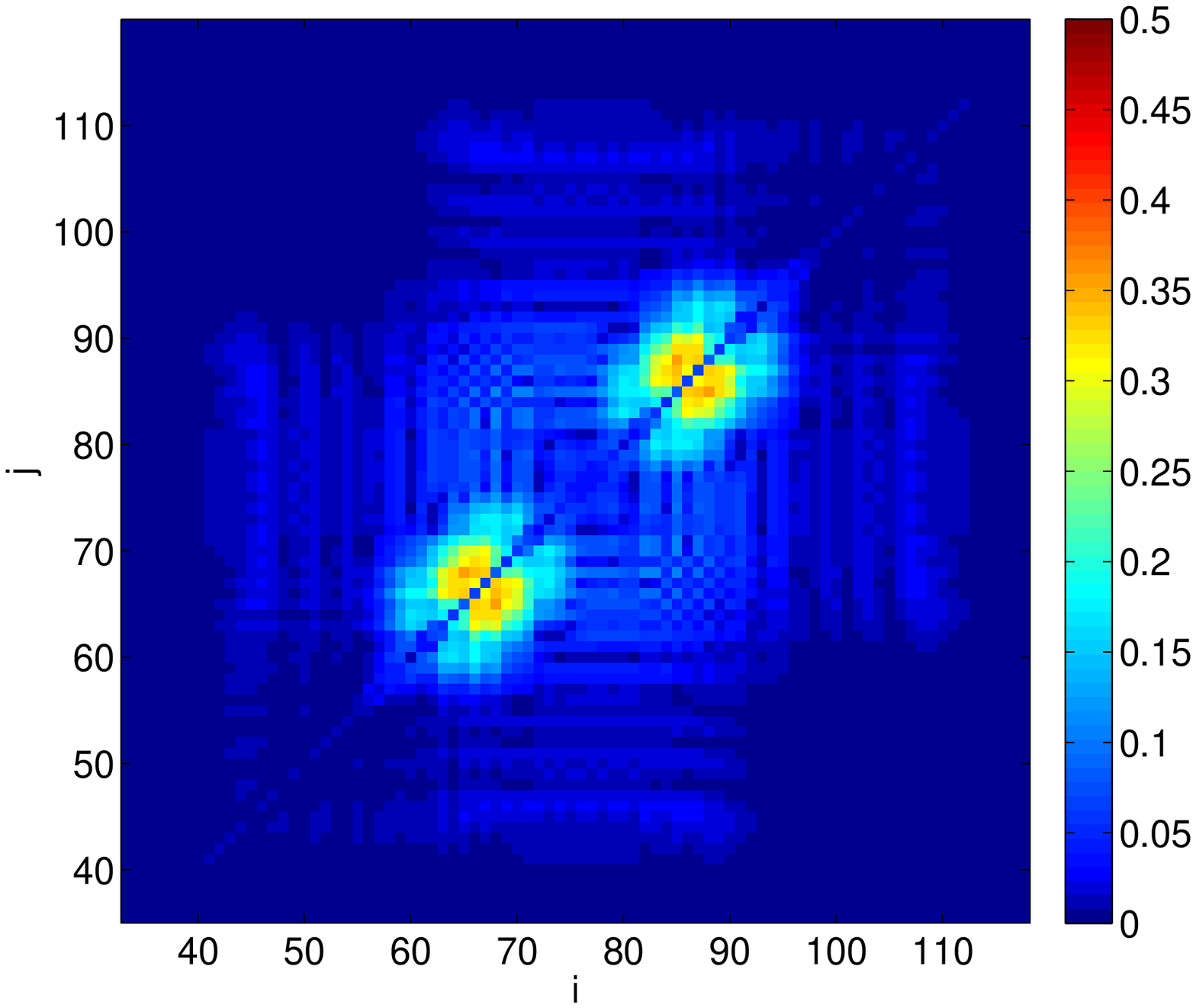} 
\includegraphics[width=0.3\textwidth]{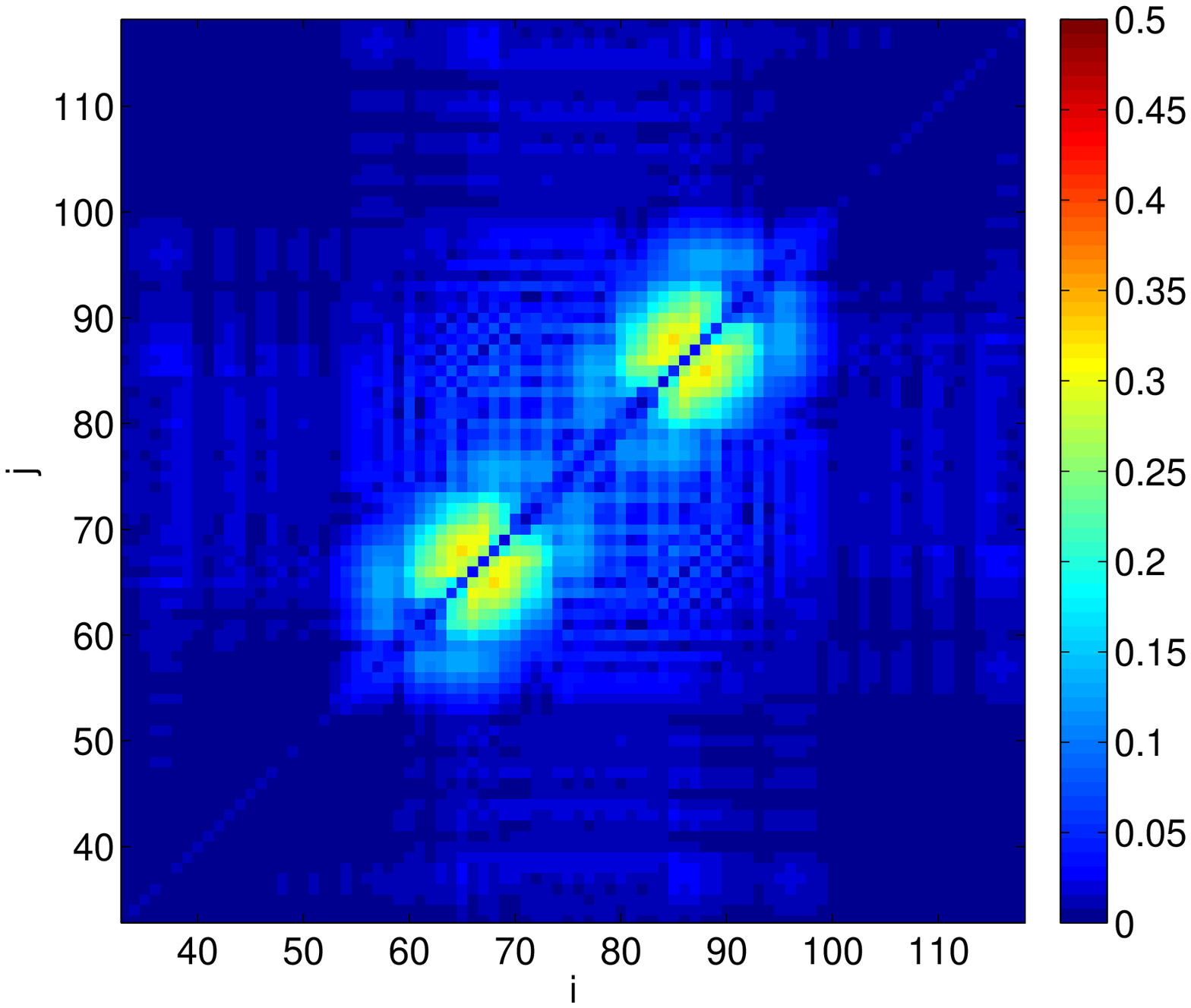}
\caption { (color online) Left: The same as figure \ref{fig:ccCorrA} 
but for $t = 12 \frac{1}{J}$ (left) and $t = 16 \frac{1}{J}$ (right).}
\label{fig:ccCorrB} 
\end{figure}

The square root of the correlator $D_{i, j} (t)$ is plotted in
Figs. \ref{fig:ccCorrA} and \ref{fig:ccCorrB} for $|U| = 5.0$ at times
$ 4 \frac{1}{J}, 8 \frac{1}{J}, 12 \frac{1}{J}, 16 \frac{1}{J}$. At
time $t = 0 \frac{1}{J}$ the correlator is zero everywhere.  We plot
the square root instead of the actual value to see better the low
value regions of the correlator. While keeping $\Gamma=150$ during the
time evolution in order to minimize the truncation error propagation,
for the calculation of the correlators we have truncated the Schmidt
number to $\Gamma = 80$ due to heavy numerical cost of the calculation,
since for the evaluation of an observable at a given time the
introduced error is negligible. Our dimer picture is
supported by the fact that most of the correlations arise between
sites around the edges of the cloud. In addition to this, correlations
arise between the wavefronts expanding in the initially free sites
$i<E_L$ and $i>E_R$ and the whole initial cloud, corresponding to the
stripes appearing in Figs. \ref{fig:ccCorrA} and \ref{fig:ccCorrB}.
A similar description can be applied to the holes
propagating towards the center of the cloud.  This behavior reflects
the intrinsically collective nature of the excitations in 1D
systems. Our picture of a ballistic particle leaving the cloud
and a hole propagating towards the centre is
complemented by this observation, which leads to the buildup of
correlations between the propagating particles and the entire
cloud.

% \bibliographystyle{apsrev}
% \bibliography{BI_bib}

\end{document}